\documentclass[aps,prb,twocolumn,superscriptaddress,showpacs,floatfix]{revtex4}

\usepackage{graphicx}
\usepackage[squaren,thinspace,textstyle]{SIunits}
\addunit{\evolt}{e\volt}

\newcommand{\iside}{\ensuremath{I_{\text{side}}{}}}
\newcommand{\iC}{\ensuremath{I_{\text{C}}}}
\newcommand{\iE}{\ensuremath{I_{\text{E}}}}
\newcommand{\kF}{\ensuremath{k_{\text{F}}}}
\newcommand{\vBC}{\ensuremath{V_{\text{BC}}}}
\newcommand{\vBE}{\ensuremath{V_{\text{BE}}}}
\newcommand{\vE}{\ensuremath{V_{\text{E}}}}
\newcommand{\rc}{\ensuremath{R_{\text{c}}}}
\newcommand{\omegac}{\ensuremath{\omega_{\text{c}}}}
\newcommand{\meff}{\ensuremath{m_{\text{eff}}}}
\newcommand{\lc}{\ensuremath{l_{\text{c}}}}
\newcommand{\lec}{\ensuremath{L_{\text{EC}}}}
\newcommand{\lee}{\ensuremath{l_{\text{e-e}}}}
\newcommand{\lph}{\ensuremath{l_{\text{e-ph}}}}
\newcommand{\tph}{\ensuremath{\tau_{\text{e-ph}}}}
\newcommand{\nph}{\ensuremath{n_{\text{ph}}}}
\newcommand{\eF}{\ensuremath{E_{\text{F}}}}
\newcommand{\eph}{\ensuremath{E_{\text{ph}}}}
\newcommand{\eBC}{\ensuremath{E_{\text{BC}}}}
\newcommand{\ekin}{\ensuremath{E_{\text{kin}}}}
\newcommand{\excess}{\ensuremath{E_{\text{kin}}-E_{\text{F}}}}
\newcommand{\eve}{\ensuremath{|eV_{\text{E}}|}}

\usepackage{color}
\definecolor{darkgreen}{rgb}{0,.55,0}

\usepackage[normalem]{ulem}

\begin{document}

\title{Relaxation of hot electrons in a degenerate two-dimensional electron system: transition to one-dimensional scattering}


% \author{}
%\thanks{}
% \affiliation{}

\author{D. Taubert}
\affiliation{Center for NanoScience and Fakult\"at f\"ur Physik,
Ludwig-Maximilians-Universit\"at,
Geschwister-Scholl-Platz 1, 80539 M\"unchen, Germany}

\author{C. Tomaras}
\affiliation{Center for NanoScience and Fakult\"at f\"ur Physik,
Ludwig-Maximilians-Universit\"at,
Geschwister-Scholl-Platz 1, 80539 M\"unchen, Germany}

\author{G. J. Schinner}
\affiliation{Center for NanoScience and Fakult\"at f\"ur Physik,
Ludwig-Maximilians-Universit\"at,
Geschwister-Scholl-Platz 1, 80539 M\"unchen, Germany}

\author{H. P. Tranitz}
\affiliation{Institut f\"ur Experimentelle Physik,Universit\"at Regensburg,
93040 Regensburg, Germany}

\author{W. Wegscheider}
\affiliation{Solid State Physics Laboratory, ETH Zurich, 8093 Zurich,
Switzerland}

\author{S. Kehrein}
\affiliation{Center for NanoScience and Fakult\"at f\"ur Physik,
Ludwig-Maximilians-Universit\"at,
Geschwister-Scholl-Platz 1, 80539 M\"unchen, Germany}

\author{S. Ludwig}
\affiliation{Center for NanoScience and Fakult\"at f\"ur Physik,
Ludwig-Maximilians-Universit\"at,
Geschwister-Scholl-Platz 1, 80539 M\"unchen, Germany}

\date{\today}

\begin{abstract}

The energy relaxation channels of hot electrons far from thermal equilibrium in a
degenerate two-dimensional electron system are investigated in transport experiments in a
mesoscopic three-terminal device. We observe a transition from two dimensions at zero
magnetic field to quasi--one-dimensional scattering of the hot electrons in a strong
magnetic field. In the two-dimensional case electron-electron scattering is the dominant
relaxation mechanism, while the emission of optical phonons becomes more and more
important as the magnetic field is increased. The observation of up to 11 optical phonons
emitted per hot electron allows us to determine the onset energy of LO phonons in GaAs at
cryogenic temperatures with a high precision,  $\eph=36.0\pm0.1\,$meV. Numerical
calculations of electron-electron scattering and the emission of optical phonons underline
our interpretation in terms of a transition to one-dimensional dynamics.
\end{abstract}

% If in two-column mode, this environment will change to single-column
% format so that long equations can be displayed. Use
%\begin{widetext}
%\end{widetext}

%\pacs{73.23.--b, 67.10.Jn, 73.50.Gr}	- verwendet im hotelectrons-Paper

%63.20.-e 		Phonons in crystal lattices
  %63.20.dd 		Measurements
  %63.20.dh 		Fitted theory
  %63.20.dk 		First-principles theory
  %63.20.kd 		Phonon-electron interactions
%67.10.Jn		Transport properties and hydrodynamics
%73.23.Ad 		Ballistic transport 
%73.23.--b		Electronic transport in mesoscopic systems
%73.50.Gr		Charge carriers: generation, recombination, lifetime, trapping, mean free paths 
%73.63.-b 		Electronic transport in nanoscale materials and structures
%72.10.Di 		Scattering by phonons, magnons, and other nonlocalized excitations (THEORY)

\maketitle

\section{introduction}

Non-equilibrium phenomena on the nanoscale increasingly gain interest as more and more devices are based on nanoscale electronics.
The ongoing miniaturization trend results in state-of-the-art transistors, used for information processing, in which only a small
number of electrons is moving through a conducting channel at a given time. Semiconductor-based quantum information
processing relies on the coherent dynamics in nanostructures. For both
classical and, in particular, quantum circuits the detection of information relies on electronic signals strong enough to be
measurable. An inevitable consequence are interactions in nonequilibrium giving rise to quantum noise and back-action. Our
experiments aim at understanding the underlying physics of interacting nanoscale circuits where only few electrons far from
thermal equilibrium carry the information.

We study in transport experiments at low temperatures the simplest case of individual nonequilibrium electrons.
After their injection these ``hot'' electrons move at first ballistically with well-defined, high kinetic energy before they relax
in an
otherwise degenerate Fermi liquid | a cold two-dimensional electron system (2DES). We are specifically interested in the case of a
high mobility 2DES since here electrons near the Fermi edge have a momentum mean-free path $l_\mathrm m$ of several micrometers
and 
therefore move ballistically through the mesoscopic device. The scattering length of the hot electrons, however, at
first strongly decreases as a function of kinetic energy as the phase space for scattering processes grows. For larger
kinetic energies the electron-electron scattering length increases again because of the high
velocity and short
interaction times.

In a magnetic field perpendicular to the plane of the 2DES, the Lorentz force tends to guide hot electrons to move
along the edges of the conducting mesa of the 2DES. In this article we focus on the transition between two-dimensional
scattering in a low
magnetic field to quasi--one-dimensional scattering in the quantum limit of edge channel transport of the hot electrons.
``Quasi--one-dimensional'' refers to a situation where the lateral width of the effective transport channel
still exceeds the magnetic length. 

Relevant energy-loss mechanisms of hot electrons in a 2DES are the emission of plasmons, acoustic or optical phonons, or
scattering with ``cold'' equilibrium electrons in the degenerate Fermi sea. The radiation of photons is strongly suppressed because of the momentum missmatch. The emission of plasmons by hot electrons is
particularly hard to capture in pure transport experiments so we will
address plasmons only briefly from the theoretical point of view in this article. The interaction between hot electrons and
acoustic phonons constitutes a minor contribution to the relaxation of hot electrons in our nonequilibrium experiments.
It
therefore won't be a focus here, but detailed investigations in zero magnetic field have already been conducted on comparable
samples by using a novel phonon spectroscopy technique.\cite{georg} It is worth mentioning, however, that acoustic
phonons play a
major role in interactions between \emph{electrically separated} nanostructures in nonequilibrium
without\cite{Khrapai2006,Khrapai2007,Khrapai2008,Gasser2009,
Harbusch2010} and with\cite{Prokudina2010} a perpendicular magnetic field applied.

The emission of optical phonons has been studied in zero magnetic field in various types of experiments since the 1960's. It
usually shows up as a very weak oscillatory signal as a function of the kinetic energy of hot
electrons, with an oscillation frequency equal to the phonon energy. It has been observed in several materials
including GaAs in photoconductivity,\cite{stocker_InSb,stocker_InSb_theory,Barker_montecarlo,nahory_photoconductivity}
in Raman scattering experiments\cite{optics_ps_1} as well as in pure transport experiments in which electrons tunnel
vertically
between layers of a
heterostructure.\cite{katayama_oldphonon,cavenett_oldphonon,hickmott_magnetophonon,lu_franckhertz,heiblum_phonon} Later the
emission of optical phonons has also been observed in lateral devices defined in GaAs-based
heterostructures.\cite{sivan_phonon,kaya}

The scattering of hot electrons in zero magnetic field with a cold 2DES has been experimentally investigated in lateral three-terminal devices
where the three regions have been separated by electrostatic barriers.\cite{sivan1989,palevski,kaya,hotelectrons,icps} It has been
demonstrated that the electron-electron interaction in such mesoscopic three-terminal devices gives rise to effects such as a
``negative resistance''\cite{kaya} and can be used for avalanche amplification of a current of hot injected
electrons.\cite{hotelectrons,icps} 

Building on Refs.\ \onlinecite{hotelectrons} and \onlinecite{icps} here we find that the transition from two-dimensional to one-dimensional scattering, as a perpendicular magnetic field is
increased, goes along with a change in the importance of electron-electron scattering versus the emission of optical
phonons. These two processes dominate the scattering dynamics of hot electrons in a degenerate two- or one-dimensional
electron system in the whole range of available magnetic fields. At low fields hot electrons relax mainly
via electron-electron scattering while the emission of optical phonons becomes more and more important at high fields. Still,
amplification of the injected electron current based on electron-electron scattering occurs even in a large perpendicular magnetic field. Our
measurements indicate that the scattering time between hot electrons and optical phonons strongly decreases as a function of the
perpendicular magnetic field while the electron-electron scattering time increases. Our experimental results are backed
up by
numerical calculations of the electron-electron scattering time and the emission rate of optical phonons as a function of
the perpendicular magnetic field. Finally, changing the magnetic field direction reveals a contribution of
hopping transport via high energy localized bulk states of the 2DES where a strong perpendicular magnetic field hinders the screening of disorder.

%\tableofcontents

\section{Sample and setup}
\label{sec-sample}

Our sample is based on a GaAs/AlGaAs heterostructure with a 2DES \unit{90}{\nano\meter} below the surface. The charge carrier
density is $n_s= \unit{2.7 $\times$ \power{10}{11}} {\mathrm{cm}^{-2}}$ which corresponds to a Fermi energy of $\eF =
\unit{9.7}{\milli\evolt}$ (determined at a temperature of $T=\unit{260}{\milli\kelvin}$). Its electron mobility
(measured at the temperature
$T=1\,{\kelvin}$) is $\mu = \unit{1.4 $\times$ \power{10}{6}}{\centi\meter\squared\per(\volt\second)}$. The resulting
equilibrium momentum mean-free path is $l_\mathrm m\simeq12\,\mu$m, an order of magnitude longer than the relevant
distances in
our sample.

\begin{figure}[ht]
\includegraphics[width=\columnwidth]{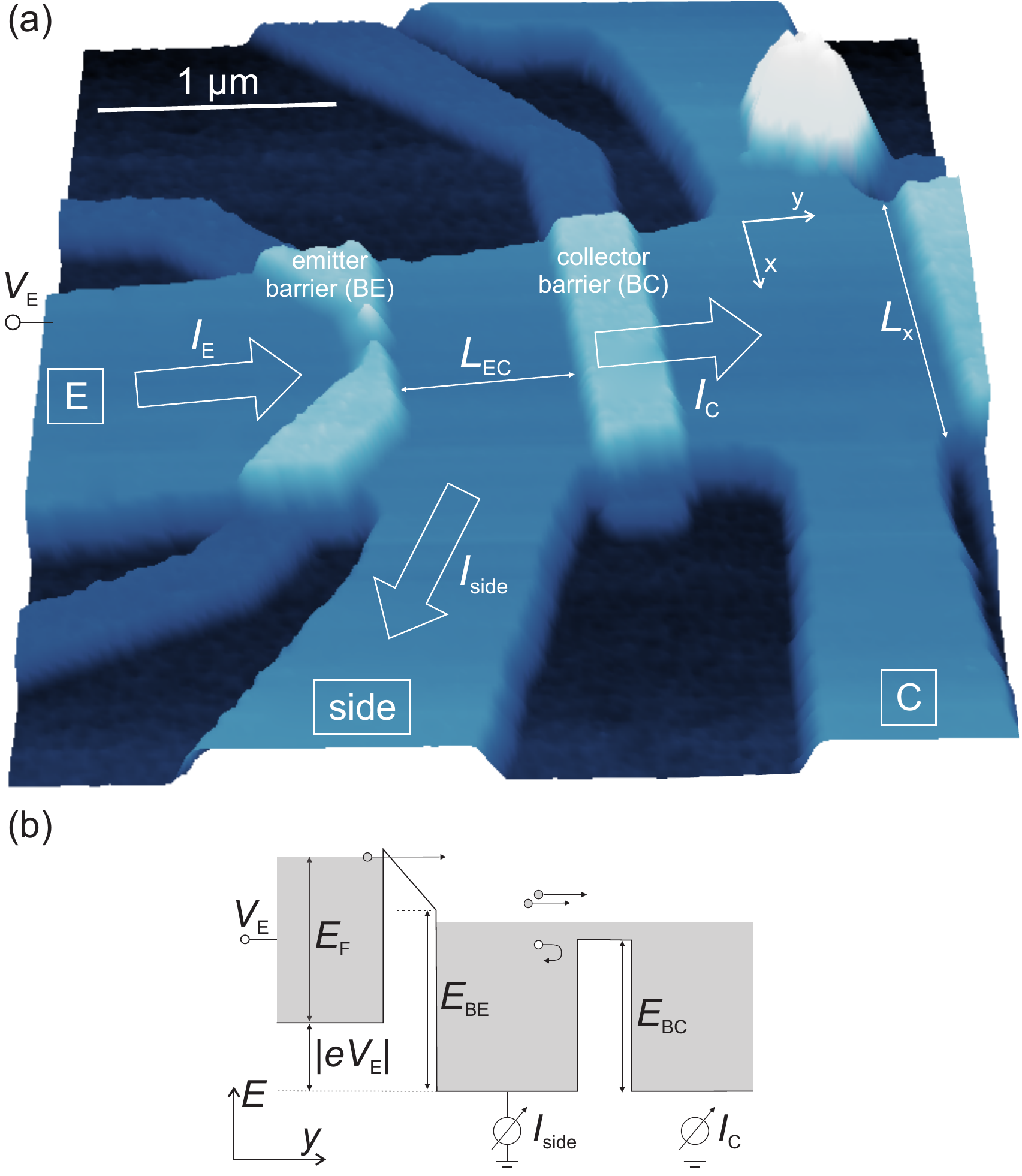}%
\caption{\label{sample}(Color online) (a) Atomic force micrograph of the sample. The elevated
areas still contain a 2DES below the surface, which is removed by wet-etching elsewhere. In addition, metal electrons
are
visible on top of the Hall bar marked as emitter (BE) and collector (BC). (b) Model that explains the avalanche
amplification effect which is based on scattering of hot electrons with the degenerate Fermi sea
of the 2DES at low temperatures.}
\end{figure}

A Hall-bar--like mesa, visible as elevated area in the scanning electron micrograph in Fig.\ \ref{sample}(a),
was created by electron-beam lithography and wet etching. The device has several ohmic contacts, three of which are used in the
experiments as the emitter ``E'', the collector ``C'' and an additional ``side'' contact. Seven metal electrodes [not
all of them
are visible in Fig.\ \ref{sample}(a)] have been fabricated by electron-beam lithography and evaporation, and are used to
create
electrostatic barriers by applying gate voltages. The two barriers used here are named emitter barrier BE and collector barrier
BC. The latter has been designed as a broad (300\,nm gate width) classical barrier whereas the former is actually a quantum point
contact. Note that the exact nature of the emitter is not important in the experiments shown here (similar measurements with a
broad barrier as emitter yield comparable results). The sample was measured in a $^3$He cryostat at an electron temperature
of $T\sim260\,$mK.

Unless stated otherwise a negative dc voltage is applied to the emitter contact
while the dc currents \iside\ and \iC\ flowing into the grounded collector and side contacts are measured with low noise current
amplifiers. The electric potentials on all other ohmic contacts of the nanostructure are left floating. Great care was taken to
tune the voltage offsets at the inputs of both current amplifiers to zero and hence avoid a superimposed  current between side
contact and collector. The emitter current \iE\ is derived via Kirchhoff's current law, $\iE = \iside + \iC$, where we define the
three currents to be positive if electrons flow into the sample from the emitter and leave the sample at the side contact and the
collector [compare arrows in Fig.\ \ref{sample}(a)]. This scenario would be expected in the limit of diffusive transport
for which the sample can be described as a network of Ohmic resistors.

\section{Avalanche amplification}
\label{sec-review}

Electron-electron scattering can cause major deviations from this ohmic case, as we have already discussed in detail in Refs.\
\onlinecite{hotelectrons} and \onlinecite{icps}. The following paragraph gives a brief overview of their main results. We keep the
emitter barrier BE nearly pinched off and hence almost the entire voltage \vE\ applied to the emitter drops across BE.
All
electrons injected via BE into the central region of the sample then have an excess kinetic energy $\excess$ close to $\eve$
because of the barrier's transmission probability that depends exponentially on energy. These \emph{hot} electrons then excite via
electron-electron scattering \emph{cold} electrons which leave behind unoccupied states in the
otherwise degenerate Fermi sea. If we name these unoccupied states in the conduction band ``holes'' (not to be confused with
valence band holes) the scattering process far from thermal equilibrium, schematically depicted in Fig.\
\ref{sample}(b), can be
interpreted as excitation of electron-hole pairs. If BC is configured such that excited electrons can pass but most holes are
reflected\footnote{For charge separation to work effectively, BC has to be slightly below the Fermi energy. See Ref.\
\onlinecite{icps} for a detailed explanation.}, the positively charged holes will reside
between BE and BC until they will
eventually be neutralized by electrons drawn in from the side contact. In its extreme, this can cause $\iE<0$ and consequently
$\iC>\iE$, where more electrons leave the device at the collector compared to the number of injected electrons. This is the case of
avalanche amplification of the injected current.

\begin{figure}[ht]
\includegraphics[width=\columnwidth]{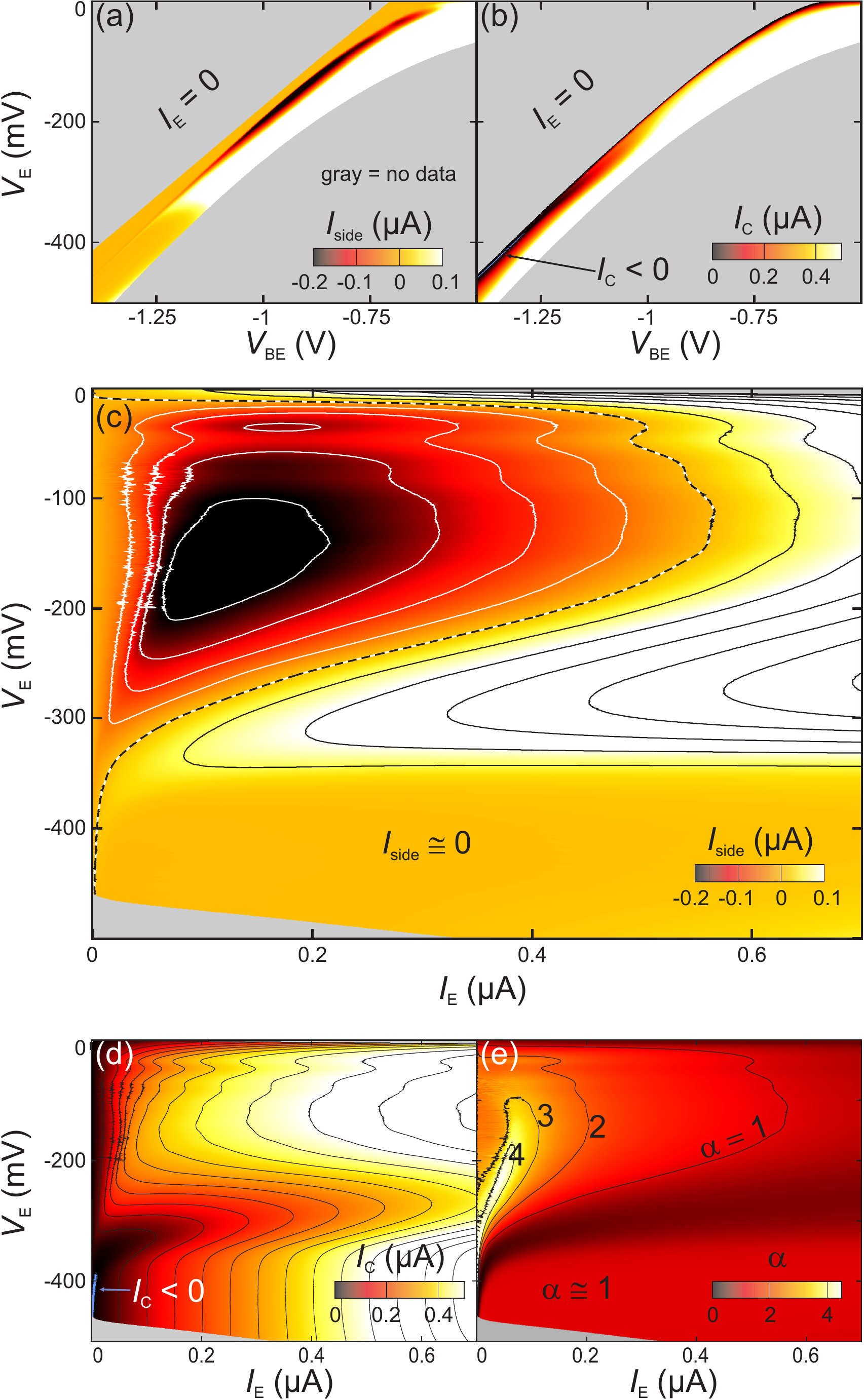}%
 \caption{\label{current_color_zerofield} (Color online) Measurements at constant $\vBC =
\unit{+45}{\milli\volt}$ ($\eBC = \eF - \unit{1.5}{\milli\evolt}$; see Ref.\ \onlinecite{icps} for details on
determining the barrier height in units of energy).  (a,b) \iside\ and \iC\ as a function of \vBE\ and \vE. (c) \iside\ as in (a), but plotted as a function of the injected
current \iE\ instead of \vBE. An extended region at the bottom (yellow) has $\iside\simeq0$ (see Ref.\
\onlinecite{icps} for details). Contour lines are spaced by \unit{50}{\nano\ampere}; white for $\iside<0$, black for
$\iside>0$ and dashed for $\iside=0$. (d) Collector current,
same type of plot as in (c). A very small area of $\iC <0$ is marked in the bottom left corner. (e) Current transfer ratio $\alpha =
\iC/\iE$, same axes. Contour lines at $\alpha=1,2,3,4$; the extended range of constant current at the bottom 
corresponds to $\alpha \simeq 1$.}
\end{figure}

Fig.\ \ref{current_color_zerofield}(a) and \ref{current_color_zerofield}(b)
depict \iside\ and \iC\ measured as the gate voltage \vBE\ (x-axis), defining the emitter
barrier BE, and the bias voltage \vE\ (y-axis) have been varied. Data are only taken in
the regime of BE nearly pinched off, namely within the roughly diagonal stripe of the
graph with variable color. The upper left regions of the two plots (gray) are
characterized by $\iE = \iside = \iC = 0$ and have therefore not been mapped out in
detail. To prevent excessive heating of the sample the overall dissipated power has been
limited to $\iE\vE\le700\,$nW which results in no data for the lower right regions (also
gray) of Figs.\ \ref{current_color_zerofield}(a) and \ref{current_color_zerofield}(b). The
gate voltage \vBE\ controls the emitter current \iE\ while the injected electrons have an
energy close to $\ekin=\eF+\eve$. Within the measured area of Fig.\
\ref{current_color_zerofield}(a), a narrow stripe of $\iside < 0$ is visible in which
amplification of the injected current occurs \cite{hotelectrons}.

Fig.\ \ref{current_color_zerofield}(c)  shows the amplification effect in a more instructive way. The raw data from
Fig.\
\ref{current_color_zerofield}(a) is plotted as a function of the injected current \iE\ and the bias voltage \vE. The
area
of negative side current is enclosed by the dashed contour line marking $\iside=0$. Other contour lines
of constant current are white at $\iside<0$ and black at $\iside>0$. \iside\ strongly depends on the energy  of the injected
electrons. This behavior is caused by the energy dependence of the electron-electron scattering length $\lee(e\vE)$. Near its
absolute minimum at $\eve\simeq150\,$meV, \lee\ is actually smaller than the distance \lec\ between BE
and
BC.\cite{hotelectrons} In this regime multiple scattering processes lead to the excitation of a large number of electron-hole
pairs between BE and BC. If only excited electrons with $\ekin>\eF$ can escape via BC (tuned to a barrier height near the Fermi
edge) a positive charge can built up between BE and BC which results in $\iside<0$ as the side contact is grounded.

At a slightly larger energy of the injected electrons our data suggest $\lee\gtrsim L_\mathrm{EC}$\cite{hotelectrons},
meaning that scattering tends to
happen just beyond the collector barrier. In this regime, electrons can be backscattered from behind BC, and then leave the sample
at the grounded side contact. This manifests itself in a positive side current and a small --- in some cases even negative --- collector
current. The latter can be seen in Fig.\ \ref{current_color_zerofield}(b) or better \ref{current_color_zerofield}(d),
which shows
\iC\ as a function of \iE\ and \vE\ in the same fashion as for \iside\ in Fig.\ \ref{current_color_zerofield}(c). In
Figs.\ \ref{current_color_zerofield}(c) and \ref{current_color_zerofield}(d) the positive \iside\ and rather small \iC\
are
clearly visible as ridge-like structures at $\eve \sim
\unit{300}{\milli\evolt}$ (see contour lines).

For $\left|\vE \right|>350\,$mV we find $\iside\simeq0$ [Fig.\ \ref{current_color_zerofield}(a)]. This can be
interpreted in terms of \lee\ exceeding the sample dimensions so that the injected electrons pass the device without
electron-electron scattering. Most injected electrons then move ballistically into the collector
contact.\cite{hotelectrons} Fig.\ \ref{current_color_zerofield}(e) shows the current transfer ratio defined as $\alpha =
\iC / \iE$ for the same set of data
as in the other subfigures. The maximal value observed in the data
presented here is $\alpha \simeq 4.5$, although we have already reached higher amplification factors of $\alpha\simeq8$ in a
different sample. $\iside\simeq0$ in the high-energy region in Fig.\ \ref{current_color_zerofield}(a) corresponds to
$\alpha\simeq1$ and $\iC\simeq\iE$ in Figs.\  \ref{current_color_zerofield}(e) and \ref{current_color_zerofield}(d),
respectively.

Note that the highest excess kinetic energies studied here --- up to $\ekin\simeq500\,{\milli\evolt}$ --- are large
compared to the Fermi
energy of $\eF\simeq{9.7}\,{\milli\evolt}$ and also exceed the energy of optical phonons ($E_\mathrm{ph}\simeq36\,$meV in
GaAs) by far. In the data presented in
Fig.\ \ref{current_color_zerofield}(c) the emission of optical phonons is faintly visible as ``wiggles'' in the contour
lines with
extrema at $\vE=36\,$mV and $\vE=72\,$mV. Optical phonons will be discussed in detail in Sec.\ \ref{sec-phonon} in
context of a magnetic field applied perpendicularly to the plane of the 2DES. 

It should be mentioned that the energy range used for the hot electrons in our experiments also exceeds the intersubband
energy of the 2DES which is in the order of $30\,$meV. Nevertheless we do not observe any signatures of
intersubband
scattering which, therefore, seems to be inefficient compared to intrasubband scattering. Most probably the majority of
hot
electrons still occupy the lowest subband which they used to populate in the emitter contact before being injected
across BE. Moreover, the maximum energy of
$\eve=500\,$meV is even larger than the vertical confinement energy of the 2DES, which could cause scattering of hot
electrons into three-dimensional bulk states. We have not observed any signatures of transport through the bulk of
the heterostructure in our experiments, though.

\section{From two dimensions to one-dimensional electron-electron scattering in a perpendicular magnetic field}
\label{sec-transition}
\subsection{Magnetic field dependence of electron-electron scattering}
\label{sec-magnetic}

In a magnetic field $B$ perpendicular to the plane of the 2DES the Lorentz force $\vec F_\mathrm L = e \vec v \times \vec
B$ acts on electrons perpendicularly to their momentary velocity $\vec v$ and forces them to move along the edges of the
conducting mesa. In a simple classical picture ballistic electrons are again and again reflected at the edge and move along
well-defined skipping orbits with the cyclotron radius $\rc={|\vec v|}/{\omegac}$ and the cyclotron frequency
$\omegac={|eB|}/{\meff}$, where $\meff$ is the effective mass of the electrons. This classical limit has indeed often been
observed in magnetic focusing experiments close to equilibrium and in moderate magnetic fields (an example including
nonequilibrium scattering is Ref.\
\onlinecite{Williamson1990}). In a quantum-mechanical description the angular momentum quantization only allows
cyclotron radii
with $\rc\left(n\right) =\sqrt{2(n+0.5)}\,\lc$, which for $n=0$ is equal to the so-called magnetic length
$\lc=\sqrt{{\hbar}/{|eB|}}$, and where
$n=0,1,2\dots$ is the Landau level index. For the case of $\ekin \sim  \eF$ and strong quantization (small
$n$) the Landauer-B\"uttiker description of the Quantum Hall effect has often been used. This model assumes a nondissipative
one-dimensional motion of current-carrying electrons within edge channels.\cite{Buettiker1986} While in the past electrons close
to thermal equilibrium have been studied in a perpendicular magnetic field, here we are interested in hot electrons far
from
thermal equilibrium with $\ekin\gg\eF$. In the following the Landau level $n$ therefore pertains to the energy of the
hot electrons rather
than the number of filled Landau levels in a degenerate 2DES. In this nonequilibrium situation we aim at observing the transition from a
two-dimensional
electron system (large \rc\ and large $n$) to the case of one-dimensional motion (small \rc\ and small $n$) in a strong
perpendicular magnetic field. Since $\rc\propto |\vec{v}/B|$ we
expect to observe one-dimensional behavior to occur at sizable magnetic fields and not too large kinetic energies of the hot
electrons.

\begin{figure}[ht]
\includegraphics[width=\columnwidth]{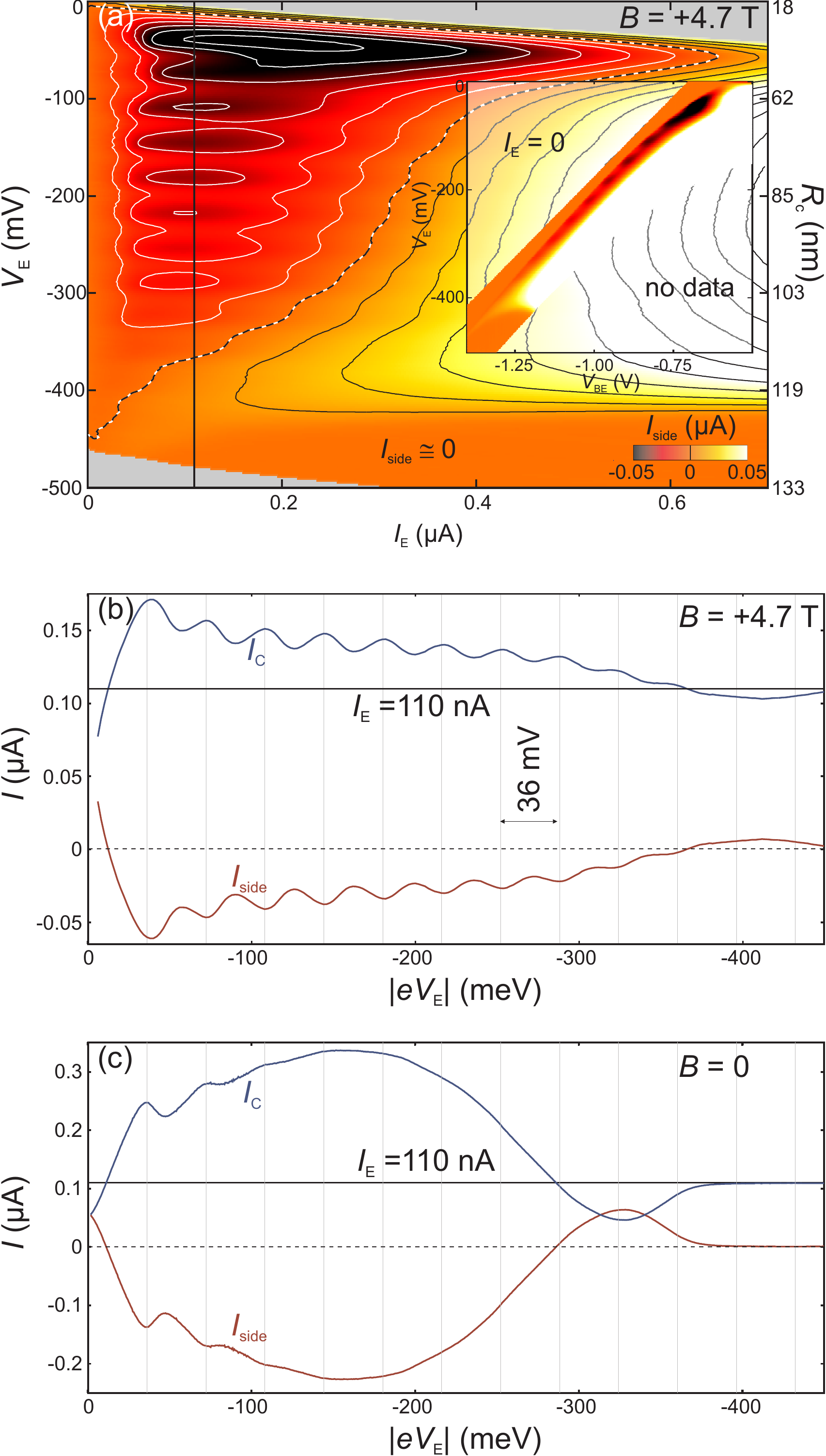}%
 \caption{\label{current_color} (Color online) (a) Side current as a function of injected current \iE\ and bias voltage 
 \vE\ in a
perpendicular magnetic field of 
$B=4.7\,$T; contour lines are spaced by \unit{10}{\nano\ampere} and colored as in Fig.\
\ref{current_color_zerofield}(c). The inset shows the raw data as a function of \vE\ and \vBC\ [as for $B=0$ in Fig.\
\ref{current_color_zerofield}(a)]. (b) 
\iside\ and \iC\ as a function of the excess energy \eve\ of the injected electrons for a constant $\iE=110\,$nA
[horizontal solid line, vertical line in Fig.\ \ref{current_color}(a)]. (c) Energy dependence of the currents as in (b)
but for $B=0$ (vertical trace of Fig.\ \ref{current_color_zerofield}(c) for $\iE=110\,$nA). Vertical lines in (b) and
(c) are spaced by $\eph=36\,$meV.
}
\end{figure}

To achieve this we have extended the previous results\cite{hotelectrons,icps} summarized above by applying a perpendicular
magnetic field. Fig.\ \ref{current_color}(a)
shows a measurement analogous to Fig.\ \ref{current_color_zerofield}(c), but in a magnetic field of
$B=4.7\,$T
which corresponds to a filling factor of $\nu\simeq1.9$ ($n=\mathrm{int}(\nu/2)=1$) for equilibrium electrons at the Fermi
energy. 
The raw data is shown as inset analogously to Fig.\ \ref{current_color_zerofield}(a). We define the sign of $B$ to be
positive
for $B$ directed upwards, i.\,e., such that electrons injected via BE into the sample are guided to the left, away from
the side contact [Fig.\ \ref{sample}(a)]. The influence of the field direction will be discussed in detail in Sec.\
\ref{sec-edges}.
Compared to the case of no magnetic field applied, at $B=4.7\,$T the emitter current is slightly reduced (by about $20\,\%$). This
is related to the conductance quantization and corresponding increase of the resistance of a 2DES in a strong perpendicular magnetic field.
However, the absolute value of the minimum
of $\iside$ and the maximum of $\alpha$ (the current transfer ratio) are much more reduced, roughly by factors of four and two,
respectively [compare Figs.\ \ref{current_color_zerofield}(c) and \ref{current_color}(a)].

A notable difference compared to zero magnetic field is that for $B=4.7\,$T \iside\ is strongly modulated with a period close
to $\delta E=36\,$meV. The oscillations of \iside\ are also mapped onto \iC\ as can best be seen in Fig.\
\ref{current_color}(b)
which plots both currents as a function of \eve\ for a fixed $\iE =110\,$nA [marked by a vertical line in Fig.\
\ref{current_color}(a) and a horizontal line in Fig.\ \ref{current_color}(b)]. These oscillations are caused by the
emission of
optical phonons with an energy $\eph\simeq36\,$meV and will be discussed in detail in Sec.\ \ref{sec-phonon}. The
analogous measurement for $B = 0$ shown in Fig.\ \ref{current_color}(c) allows a quantitative comparison near maximum
amplification. At zero magnetic field amplification clearly is much stronger (deeper minimum of \iside) while
phonon-induced oscillations are only seen at relatively small energies of $\eve\lesssim100\,$meV. These observations
imply that in our sample the electron-electron scattering rate between BE and BC decreases in a perpendicular magnetic
field while at the same time the emission of optical phonons becomes more significant.

Furthermore, the shape and current profile of the region of $\iside<0$ [framed by dashed contour lines in Figs.\
\ref{current_color_zerofield}(c) and \ref{current_color}(a)] --- corresponding to $\alpha>1$ --- changes drastically if
a
perpendicular magnetic field is applied. For relatively small injected currents \iE\ the region of $\iside<0$ extends towards
larger \eve\ in a perpendicular magnetic field. At the same time the absolute minimum of \iside (\vE), which is
relatively broad at $B=0$, shifts
towards smaller energies $\eve<100\,$meV and narrows in a magnetic field. This first (and deepest) minimum of \iside\
due to optical phonons at
$B=4.7\,$T still extends to relatively large currents up to $\iE>0.6\,\mu$A. With increasing \iE\ the main minimum shifts towards
larger \eve\ while the phonon-induced oscillations are constant in \eve. We conclude that in a large perpendicular
magnetic field the dependence of the amplification effect on the energy $\left| e \vE \right|$ is strongly altered by
the emission of optical phonons
while its dependence on the dissipated power, $P=\iE\vE$, is still dominated by electron-electron scattering .

Could the observed change in the amplification effect then be related to a transition of scattering of hot electrons in two
dimensions to one-dimensional scattering at finite $B$? A first answer to this question can be given by a comparison of the
screening length $\kF^{-1}\simeq 8\,$nm of the 2DES at $B=0$ with the width of the current-carrying channel at $B=4.7\,$T,
approximately given by the cyclotron radius in the quantum Hall regime. At $B=4.7\,$T we find $\rc(n = 1)=\lc\simeq12\,$nm, which
is in the same order of magnitude as $\kF^{-1}$. 
The main minimum of \iside\ in Fig.\ \ref{current_color}(a) at relatively large currents where the
phonon-induced oscillations are
weak occurs at energies in the order of $\eve \sim \unit{50}{\milli\evolt}$. This corresponds to 
 $\rc\ge45\,$nm (and $n\simeq7$) which is considerably larger than the two-dimensional screening
length, $\kF^{-1}\simeq 8\,$nm.
Hence we do not expect our data to resolve the complete transition from two-dimensional to fully one-dimensional
scattering of hot electrons. The comparison of the length scales suggest that our data can be interpreted as a
signature of quasi--one-dimensional scattering.

\subsection{Hot electron statistics}
\label{sec-statistics}

We start a more detailed discussion by highlighting some numbers relevant for the scattering dynamics of the hot
electrons. In the next two sections we will then present theoretical
considerations pertaining to scattering of hot electrons with a degenerate 2DES in a
perpendicular magnetic field. The area occupied by 2DES between BE and BC
measures slightly more than a square micrometer and therefore contais about 3000
electrons. The potential in this area is increased in a regime of avalanche amplification
by up to $\delta V\sim 1\,$mV (see Ref.\ \onlinecite{hotelectrons} for details on the
setup used to obtain this number) which corresponds to a reduction of the resident
electrons by roughly 10\,\%. In the following we restrict the discussion to effects at
considerably larger energy scales.

For simplicity we now assume that the injected electrons move ballistically from BE to BC
across the distance of $L_\mathrm{EC}\simeq700\,$nm. Then the average number of hot electrons traveling simultaneously
between
BE and BC is $\left<N\right>=\left|\frac{\iE L_\mathrm{EC}}{e}\right|\,\sqrt{\frac{\meff}{2(\eF+\eve)}}$. We find a maximum of
$\left<N\right>\simeq10$ for our measurements at $\iE=0.6\,\mu$A and $\vE=-10\,$meV. However, amplification
($\iside<0$) occurs at
$\left<N\right>\lesssim1$ and most of the following discussions apply to this case for which we can neglect direct Coulomb
interaction between hot electrons.

\subsection{The role of plasmons}
\label{sec-plasmons}

Scattering of hot electrons with a three-dimensional degenerate electron system is discussed in standard textbooks on
Landau
Fermi-liquid theory.\cite{Pin,Bay,Gui} In three dimensions an energy-dependent quasiparticle relaxation is usually
considered that can be divided into two regimes. On the one hand, the relaxation of excited electrons at
rather low energies with momenta $k\ll 2\kF$ is dominated by particle-hole excitations. On the other hand, at rather high energies
(in the order of several electron volts in typical metals) beyond the plasma frequency the emission of plasmons becomes
important. In two dimensions this clear separation into two regimes breaks down at
relatively high electron densities for which the Fermi energy \eF\ by far exceeds the mean mutual Coulomb energy
$\left<E_\mathrm C\right>$ between conduction band electrons (and the Thomas-Fermi wavelength exceeds the
inter-particle distance): $r_\mathrm
s\equiv\frac{\left<E_\mathrm C\right>}{\eF}\propto 1/\sqrt{n_\mathrm s}$ fulfills $r_\mathrm s\ll 1$.\cite{Gui1,Fet} Our
sample resides with $r_\mathrm s\simeq0.6$ in an intermediate regime and we expect that the relaxation of hot electrons with
sizable excess energies via the emission of plasmons should play a role at zero magnetic field. Nevertheless, our transport
measurements have proven to be relatively insensitive to the emission of plasmons and we have not been able to identify
traces of plasmon excitations in our present data.

If, in addition, a perpendicular magnetic field $B$ is applied, the two-dimensional bulk
electrons become rigid against perturbations at energies low compared to the Landau-level
separation $\hbar\omegac$. For hydrodynamic plasmons, i.\,e., in
the limit of very small momenta $q\rightarrow0$, the only difference between the two cases of $B>0$  and $B=0$ is this
finite gap $\hbar\omega_c$.  \cite{Ting1,Ting2,Sarm1,Chiu,Gree,Gass} However, at
sufficiently high magnetic field, the transport properties of a disordered 2DES are determined by the one-dimensional
edge channels, hence, the bulk magneto-plasmons would not be visible in a transport measurement. In a one-dimensional
system,
interactions are strong and the electron system can be described as a Luttinger liquid rather than a Landau
Fermi liquid. In the low-energy regime the relaxation is dominated by dephasing \cite{LeHur,Gia}. For technical
reasons, investigations of the electron-plasmon interaction in the Luttinger liquid are limited to intermediate energies
corresponding to momenta $k \lesssim 2k_F$.\cite{LL1,Glaz} One
would expect that the scattering of a single hot electron with the degenerate electron systems displays a
transition from
Landau-Fermi to Luttinger liquid behaviour as the external magnetic field is increased. This
feature is absent in the present experiment, which can be understood by considering the width of the one-dimensional
channel as described in the following. 

\subsection{Magnetic field dependent calculations}
\label{sec-calculation}

Here we calculate the electron-electron scattering length of hot electrons moving in a
quasi--one-dimensional
channel of an otherwise degenerate 2DES. The width of the quasi--one-dimensional edge channel, produced by a
perpendicular magnetic
field, is approximately given by the cyclotron radius $\rc =\sqrt{2(n+0.5)}\,\lc$. We specifically consider the realistic scenario
of weak two-dimensional scattering in which the width of the channel exceeds the screening length of the 2DES by far,
$\rc\gg\kF^{-1}$.

Edge channels form in a perpendicular magnetic field because the physical edges of the Hall bar represent boundaries which bend
the Landau levels upwards in energy. Formally, the degeneracy of the electrons in each Landau level is lifted near the
edges
according to
$E_n=E_n(p_y)$, where $p_y$ denotes the momentum component of a hot electron parallel
to the edge of the 2DES. A good approximation for etched edges is a hard-wall confinement of the 2DES, which allows an
exact
calculation of the dispersion\cite{qt}
\begin{eqnarray}\label{eq:dispersion}
E_n(p_y)=\left(n+\frac{1}{2}\right)\hbar\omegac+\frac{p^2_y-p^2_{\rm{min}}}{2l^2_\mathrm
c}\,\Theta\left(\frac{p^2_y-p^2_{\rm{min}}}{2l^2_\mathrm c}\right)\,.
\end{eqnarray}
The resulting single-electron wavefunctions resemble those of the free 2DES, $\Psi_n(p_y)\sim H_n(x-l^2_c p_y)e^{ip_y
y}$, with
$H_n$ being the $n$-th Hermite function. 
The electrons within the quasi--one-dimensional channel interact via the two-dimensional Coulomb potential 
\begin{eqnarray}
&\widehat{V}_{\mathrm{2D}}=\sum\limits_{\vec{q},\vec{k},\vec{p}}v(|q|)\sum\limits_{nmn'm'}|J_{mn}(q_{y},q_{x},0)|^{2}
&\nonumber\\
&\times c_{\vec{k}+\vec{q}|m,\sigma}^{\dagger}c_{\vec{p}-\vec{q}|n,\bar{\sigma}}^{\dagger}c_{\vec{p}|m',\bar{\sigma}}c_{\vec{k}|n',\sigma}\,,&
\end{eqnarray}
 where $c_{\vec{k}|m,\sigma}^{\dagger}$ creates an electron with momentum $\vec{k}$ in the $n$-th Landau level, $v(|q|)$ is the ordinary
two-dimensional Coulomb matrix element with the exchange momentum $\vec{q}$, and
$J_{mn}(q_x,q_y,0)$ describes the interaction between electrons in the Landau levels $n$ and $m$. Further details have
been discussed in Ref.\ \onlinecite{Gass}. In the limit of hydrodynamic transitions,
i.\,e.\ $q\rightarrow 0$, Eq.\ (2) 
reduces to the ordinary plane-wave Coulomb interaction. Giuliani \emph{et al.}\ calculated the electron-electron
scattering rate at $B=0$ \cite{Gui1}. It is straightforward to write down the
analog in presence of a perpendicular magnetic field 
\begin{eqnarray}\nonumber
&\frac{1}{\tau_m(p_{y})}=\sum\limits_{\vec{q},m,n}[\frac{v(|\vec{q}|)|J_{mn}(q_{y},q_{x},0)|^{2}}{
\pi\epsilon_{\mathrm{RPA}}(q,E_{m}(p_{y})-E_n(p_{y}+q_{y}))}]^{''}&\\
&\times(1-f_n(p_{y}+q_{y}))\,,&
\end{eqnarray}
by summing over those exchange
momenta $\vec{q}$ which are related to electron-electron interaction within the quasi--one-dimensional channel ($|q|\sim
2\pi n/l_c$). 
  Here the dielectric
susceptibility $\chi_0=(1-\epsilon_{RPA})/v(|\vec{q}|)$ is given by
\begin{eqnarray}\nonumber
&\chi_{0}(q,\omega)=\sum\limits_{mn ,k}|J_{mn}(q_{y},q_{x},0)|^{2}&\\&\times
\frac{f_m(k)-f_n(k+q_{y})}{\omega -i0^{+}+E_m(k)-E_n(k+q_{y})}\,,&
\end{eqnarray}
and momentum sums are cut off at $p_{\rm{max}}=L_x/2l^2_c$, where $L_x$ is the width of the Hall bar (see Figure 1);
$f_n$ is the Fermi distribution in the $n$-th Landau level. For our numerical evaluation we use the
quadratic dispersion in Eq.\ (\ref{eq:dispersion}) which is the exact solution for a hard-wall confinement.
Since a clear physical separation of the edge channels at high energies seems unlikely, we assume that each in-situ injected
electron with energy $\ekin$ occupies any one of the Landau-levels fulfilling $\ekin< E(n,p_y)$ with equal probability
and
average the inverse quasi-particle lifetime over these channels.

\begin{figure}[ht]
\includegraphics[width=\columnwidth]{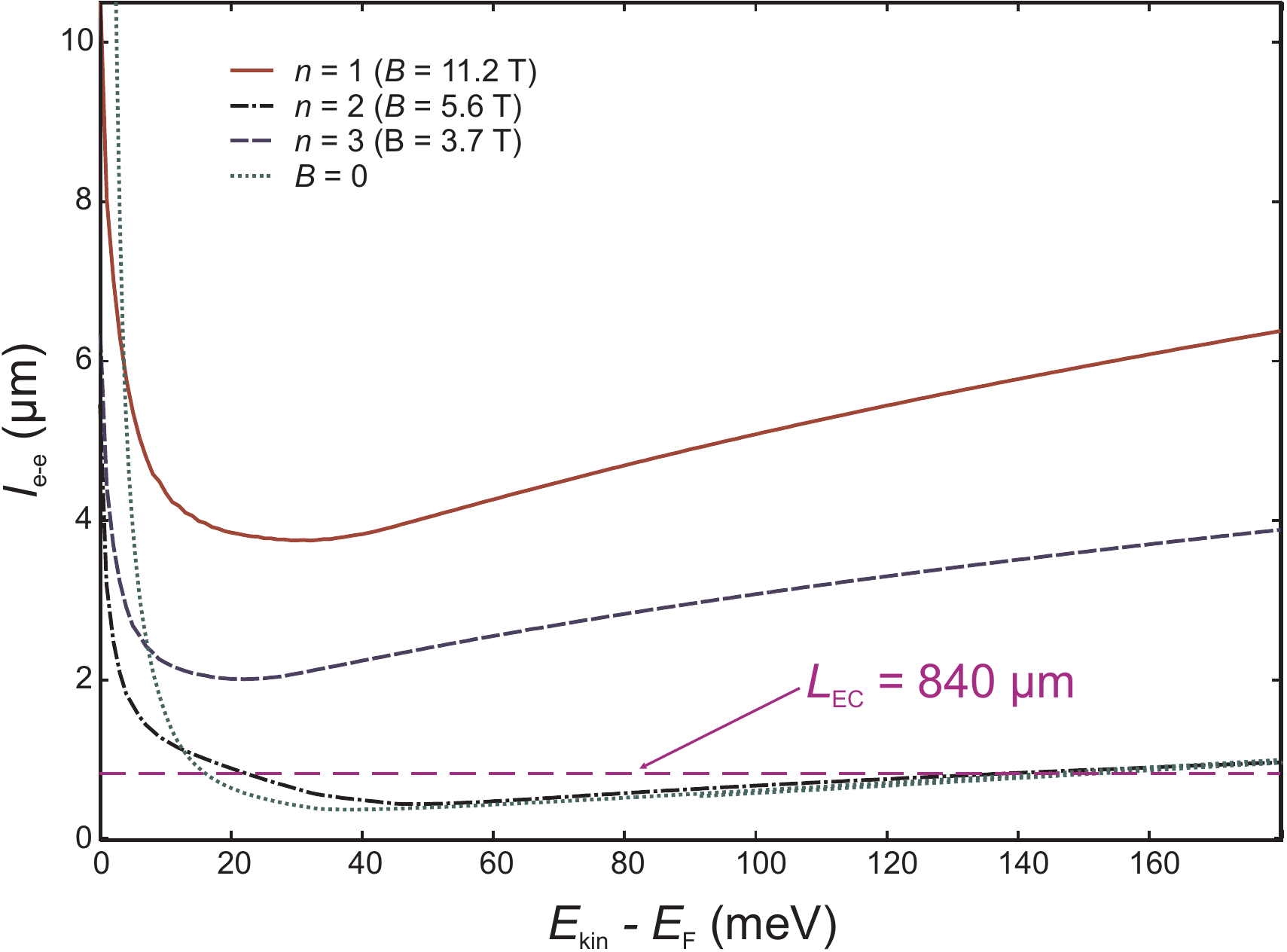}%
\caption{\label{lambda_ee} (Color online) Calculated scattering length for scattering between one hot electron with
kinetic energy \ekin\ and cold electrons of the degenerate 2DES as a function of excess kinetic energy for
different filling fractions with fully occupied Landau-levels $n=1,2,3$ and $n\rightarrow\infty$ (corresponding to
$B=0$). To compare the values of $n$ with experimental data, corresponding values of $B$ are given as well,
calculated for the charge carrier density of our sample.}
\end{figure}

In Fig.\ \ref{lambda_ee}
we show the electron-electron scattering length $\lee=|\vec v|\tau_\mathrm{ee}$ calculated with (4) for the $n=1,2,3$
Landau levels fully
occupied by the degenerate 2DES (integer filling factors $\nu=2,4,6$) as well as the $B=0$ result (corresponding to
$n\rightarrow\infty$) obtained in a previous calculation.\cite{hotelectrons} The overall shape of all curves is similar,
exhibiting a rapid decrease of \lee\ for $k\ll2\kF$ followed by a gradual increase of $\lee\propto\ekin$ in the limit $k\gg2\kF$.
This general behavior has already been discussed in detail for the case of $B=0$ in Ref.\ \onlinecite{hotelectrons}.
For $n\le 2$ the scattering length never drops
below the distance between BE and BC ($\lec \simeq840\,$nm). This explains the overall weaker amplification
(smaller $|\iside|$) which we observe at $B=4.7\,$T compared to the case of $B=0$ [compare Figs.\
\ref{current_color_zerofield}(c)
and \ref{current_color}(a)]. It can be qualitatively understood since the width of the quasi--one-dimensional channel
shrinks with
increasing magnetic field and the phase space for electron-electron scattering is hence reduced. The $n=3$ curve apparently
converges to the zero field curve at large \ekin\ while at low \ekin\ large differences remain. This behavior expresses
the
fact that at high energies forward scattering dominates which resembles scattering in a one-dimensional channel. Our calculations
suggest that the differences in the electron-electron scattering dynamics between the one-dimensional and the two-dimensional
cases disappear at high \ekin\ and $n\ge3$. In Fig.\ \ref{lambda_ee} at low energies the scattering length at small
filling fractions ($n=1,2,3$) differs from its 2DES counterpart for $B=0$ ($n\rightarrow\infty$). The origin of this
discrepancy is the altered dispersion, yielding a different asymptotic behavior of the
appropriate susceptibilities $\chi$ within the particle-hole continuum. Hence this technical discrepancy at small
energies should not be mistaken for a realistic prediction.  

In summary, the calculated scattering length \lee\ plotted in Fig.\ \ref{lambda_ee} explains several aspects of the measured
magnetic field dependence expressed in our data in Figs.\ \ref{current_color_zerofield}(c) and \ref{current_color}(a).
The increase of
\lee\ with increasing magnetic field causes the measured overall weaker amplification at finite $B$. The similarity of the
calculated $\lee(\ekin)$ curves for different values of $B$ explains the overall similar behavior (except for
phonon-induced
effects) observed at different perpendicular magnetic fields. The calculations in particular predict that amplification
should occur even at
high perpendicular magnetic fields (up to $n=1$) while the strongest amplification is expected for $B  \to 0$ $ (n > 3)$
where we find a broad minimum of $\lee < \lec$. The calculations suggest a transition towards one-dimensional
scattering to happen for $n\le3$. The increase of \lee\ with increasing magnetic field in this regime especially at large energies
suggests that the region of amplification should be restricted to lower energies. This corresponds to the narrowing of
the main minimum of \iside\ towards lower \eve\ in a finite perpendicular magnetic field.
The ballistic regime visible in Figs. \ref{current_color_zerofield}(c) and \ref{current_color}(a) as extended areas of
$\iside \simeq 0$ in the limit of large $\left| e \vE \right|$ is expected for $\lee \gg \lec$. It can therefore be
qualitatively explained by the calculated monotonic increase of \lee\ at large $\ekin - \eF$ (Fig. \ref{lambda_ee}).

We have succeeded in qualitatively explaining some of the differences between the data at zero versus finite
perpendicular magnetic
field by the comparison with our calculations of \lee. However, other features can only be understood by taking the relaxation of
hot electrons via emission of optical phonons into account. In our experiment the emission of optical phonons can be
clearly seen as periodic oscillations of the measured currents (see Fig. 3), which is the subject of the next section.

\section{Emission of optical phonons}
\label{sec-phonon}

\subsection{Correlation between electron-electron scattering and emission of optical phonons}

Electrons carrying enough excess energy $\ekin-\eF$ can lose part of their energy by emitting an optical phonon. The
electronic momenta in our experiments cover only about 10\,\% of the first Brillouin
zone. In this
range of
small wave vectors optical phonons have an almost constant dispersion relation, with an energy of $\eph(q)\simeq36\,$meV in
GaAs.\cite{waugh_dispersion,blakemore_GaAsreview,hickmott_magnetophonon,heiblum_phonon,sivan_phonon} We therefore expect
the hot electrons to emit phonons with a fixed energy $\eph\simeq36\,$meV and a wide range of momenta. In
fact, the momentum transfer is only limited by the quadratic dispersion relation of the electrons and the scattering cross-section
for the emission of optical phonons is accordingly large. Since their dispersion relation is flat, the emitted optical phonons
have almost zero velocity. Their main decay channels are reabsorption by electrons or, more efficient, the emission
of two acoustic phonons. The momenta of the resulting acoustic phonons are not confined to the plane of the 2DES;
consequently, they tend to disappear into the three-dimensional crystal and have a negligible chance to be reabsorbed
by the 2DES.

The oscillations in \iside\ and \iC\ as a function of \vE\ in Fig.\ \ref{current_color}(b) are clearly caused by the
emission of
optical phonons, as their period [see vertical lines in Fig.\ \ref{current_color}(b)] is close to the energy of the
LO phonons in
GaAs $\eph\simeq36\,$meV. The oscillations occur because the number of optical phonons that can be emitted per hot electron
increases one by one at integer multiples of $\eph$ if \eve\ is increased. At large magnetic fields we have observed up
to eleven of these oscillations which implies the
sequential emission of eleven optical phonons by the individual hot electrons. By averaging over several periods of the
oscillations we have determined the \emph{energy of the LO phonons in GaAs} at a cryogenic temperature $T\sim260\,$mK
with a high precision to be $\eph^{\text{exp}} = \unit{(36.0 $\pm$ 0.1)}{\milli\evolt}$. This result implies that
in our experiments the interaction between electrons and phonons is dominated by the bulk LO phonons. If other decay
channels, e.g., interface optical phonons which have different energies,\cite{algaas,algaas2} were also important the
observed oscillations of \iside\ would appear more irregular. The coupling between the two-dimensional electron
system in a GaAs/AlGaAs heterostructure and interface phonon modes is
expected to be reduced compared to bulk modes because of the small
probability function of the electrons at the interface.

\begin{figure}[ht]
\includegraphics[width=\columnwidth]{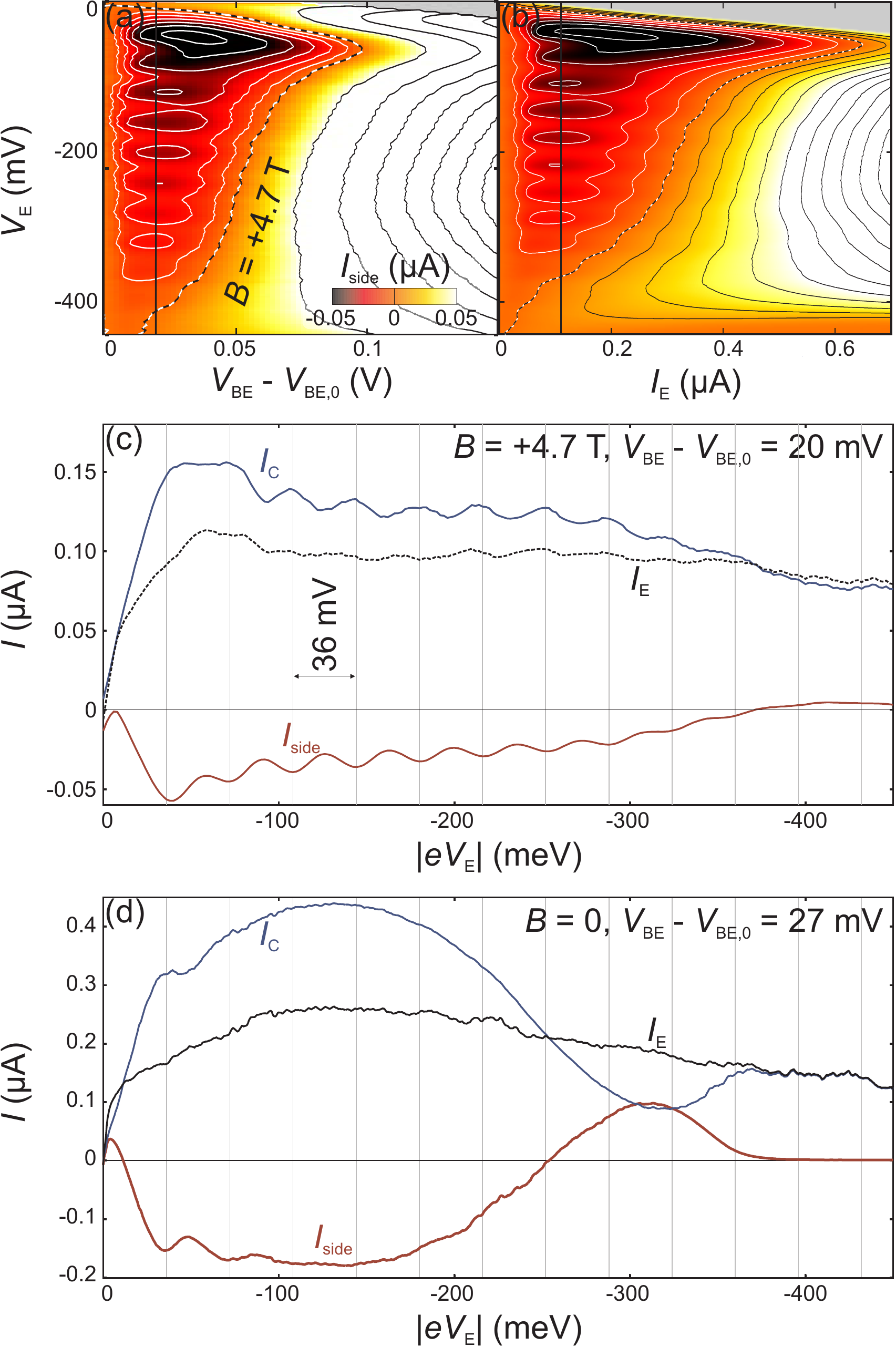}%
\caption{\label{phase}(Color online) (a) Same data as in the inset of Fig.\ \ref{current_color}(a) but as a function of
reduced gate voltage $\vBE - V_{\text{BE,0}}(\vE)$ instead of \iE\ (see main text for details). (b) Reproduction of Fig.\ \ref{current_color}(a)
to allow a direct comparison. (c) Vertical traces of (a) at $\vBE - V_{\text{BE,0}} =
\unit{20}{\milli\volt}$ as a function of \eve\ for all three currents. Integer multiples of $\eve=36\,$meV are marked by vertical lines; (d) traces
analogous to (c), but for $B = 0$ (Fig.\ \ref{current_color_zerofield}) and $\vBE -
V_{\text{BE,0}} = \unit{27}{\milli\volt}$. }
\end{figure}

\iC\ has local maxima and \iside\ local minima at integer multiples of \eph\ [vertical lines in Fig.\
\ref{current_color}(b)] which corresponds to a rigid relative phase of $\pi$ between the oscillations
of the two currents. However, in Fig.\ \ref{current_color}(b) this behavior is caused by the way the data are depicted,
namely for a fixed \iE. In fact, $\iside+\iC=\iE = \mbox{ const.}$  determines the phase between \iside\ and \iE. To
overcome this limitation we therefore replot the raw data [see inset of Fig.\
\ref{current_color}(a)] as a function of $\vBE -
V_{\text{BE,0}}(\vE)$ in Fig.\ \ref{phase}(a).
For comparison Fig.\ \ref{phase}(b) reproduces the plot for constant \iE\ already shown in Fig.\
\ref{current_color}(a). $V_{\text{BE,0}}(\vE)$ is defined as the gate voltage at which the emitter opens for a given
\vE, and
current starts to flow ($\iE>0$). Compared to the choice of a constant \iE\ this processing technique does not require a rigid
phase relation between \iside\ and \iC. The disadvantage is a small uncertainty regarding an offset of \vE\ which, however, does
not effect the relation between \iside\ and \iC. Fig.\ \ref{phase}(c) shows a vertical trace of Fig.\ \ref{phase}(a) at
$\vBE -
V_{\text{BE,0}} = \unit{20}{\milli\volt}$ and Fig.\ \ref{phase}(d) an analoguous graph for $B=0$. Interestingly,
the rigid phase
difference of $\pi$ between the phonon-induced oscillations in \iside\ and \iC\ remains (independent of $B$). This expresses a direct linkage between
\iside\ and \iC\ caused by electron-electron scattering which in itself is correlated to the emission of optical
phonons.

Regardless of the value of $B$ the emitter current, which is no longer fixed,
steeply grows at first as the emitter voltage \vE\ is increased. At $B=0$ \iE\ then develops a broad maximum at
$\eve\sim140\,$meV [Fig.\ \ref{phase}(d)]. The gradual decrease of \iE\ at  $\eve >\unit{140}{\milli\evolt}$ is caused
by
an increase of the overall resistance. This is linked to an increase of the electron-phonon scattering rate with growing
excess energy. At $B=4.7\,$T \iE\ is further reduced and, interestingly,
stays roughly constant in the regime of strong phonon-induced oscillations of \iC\ and \iside. A closer look reveals
that \iE\ is slightly modulated and decreases stepwise at integer multiples of $\eph$ [vertical lines in Fig.\
\ref{phase}(c)]. This weak modulation can be explained as an incremental increase of the overall device resistance
whenever
additional optical phonons contribute to electron scattering. It is a relative small effect since the device
resistance is dominated by the almost pinched-off emitter.

The oscillations of \iC\ and \iside\ (at finite $B$) are much stronger and cannot be explained in terms of the weak modulation of the overall resistance. Instead
they express the correlated dynamics of electron-phonon and electron-electron scattering. At relatively small excess energies
$\eve<36\,$meV the collector current \iC\ steadily increases and exceeds the emitter current $\iC>\iE$ while the side current
decreases accordingly, becoming negative. This avalanche amplification effect has been discussed in detail
in Sec.\ \ref{sec-review} and \ref{sec-transition}. In short, the increase in amplification is caused by the decrease of
\lee\ as \eve\ is
increased. At $B=4.7\,$T the amplification is reduced compared to $B=0$ while the emission of optical phonons
has become an important scattering process [oscillations in Fig.\ \ref{phase}(c)]. In the extreme case, the hot
electrons will first
emit as many optical phonons as possible and therefore end up with an energy less than \eph\ before most of the
electron-electron
scattering occurs. As expected for such a scenario the negative side current \iside(\eve) reaches its absolute minimum at
$\eve\simeq36\,$meV, just before the first optical phonon can be emitted. For higher energies \eve\ the side current then
oscillates with local minima at integer multiples of \eph\ where yet another optical phonon can be emitted.

Our calculations presented in the previous section suggest that \lee\ increases with increasing perpendicular magnetic
field as electron-electron scattering becomes weaker due to the transition from two-dimensional to
quasi--one-dimensional scattering. We will see that the emission rate of optical phonons increases under the transition
from two to one
dimensions.\cite{bockelmann_theory,tel} Our observation that the scattering with optical phonons takes over from
electron-electron scattering as the perpendicular magnetic field is increased can therefore be interpreted as indication for a transition from
two-dimensional to one-dimensional scattering.

\subsection{Magnetic field dependence of the electron-phonon scattering time}

In the following we estimate the electron-phonon scattering time \tph\ by counting phonon-induced oscillations of
\iside\ as a function of \eve. Until now we have assumed that if the electron-phonon scattering length is much smaller
than the electron-electron scattering
length ($\lph\ll\lee$), the maximum possible number of optical phonons
$\nph^\mathrm{max}=\mathrm{int}\left(\frac{\eve}{\eph}\right)$ will be emitted first. Afterwards hot electrons with the
reduced excess
energy $\excess=\eve-\nph^\mathrm{max}\eph<\eph$ will scatter with the 2DES. While this assumption seems reasonable for a rough
picture, it disregards the influence of the sample's geometry on the correlation between the emission of LO phonons and
electron-electron scattering, namely that only scattering processes happening in the region between emitter BE and collector
barrier BC contribute to the observed amplification effect. The apparent decrease of the amplitude of the phonon-induced
oscillations at large
energies \eve\ [Fig.\ \ref{phase}(c)] is related to this geometrical restriction.

Consider the length of the path $L(B)\ge \lec$ that an electron subjected to a perpendicular magnetic field travels
along the mesa edge from BE to BC. Assuming that
backscattering of the hot electrons is strongly suppressed in a large perpendicular magnetic field (the extreme case are
Landauer-B\"uttiker edge channels which only allow forward scattering), we can make an accurate statement.
Phonon-induced oscillations with undiminished amplitude will be observed if (1) $\lee \gtrsim L$ (as is the case in our
sample) and (2) $\nph(L) \le \nph^{\rm{max}}$. Here $\nph(L)$ is the number of phonons already emitted when the
electron reaches BC.

In order to roughly estimate the electron-phonon
scattering time \tph\ from our measurements in different magnetic fields we now completely neglect all other energy relaxation
processes including electron-electron scattering.
We can then express the velocity of a hot electron with initial kinetic energy $\ekin^0 = \left| e
\vE \right| + \eF$ after emission of $k$ phonons as
\begin{equation}\label{eq:vk}
v_k = \sqrt{\frac{2\left(\ekin^0-k\eph\right)}{\meff}} \,.
\end{equation}
The mean distance a hot electron travels until it has emitted \nph\ optical phonons is
\begin{equation}\label{eq:lvonn1}
l(\nph) = \sum_{k=0}^{\nph-1} v_k \tph(v_k) = \tph   \sum_{k=0}^{\nph-1} v_k \,.
\end{equation}
In the last step we assumed for simplicity that $\tph(v_k) \equiv \tph$ is independent of the energy of
the electron (as long as
$\excess>\eph$), which is justified by previous calculations suggesting only a weak energy dependence.\cite{tel}

The expectation that on average \nph\ phonons have already been emitted after the hot electron has traveled the distance
$L$ is expressed in the inequality $l(\nph)\le L<l(\nph+1)$ which leads to the approximation

\begin{eqnarray}\label{eq:tau}
L &\simeq& \frac{1}{2} \left[ l(\nph) + l(\nph+1) \right]  \nonumber\\
 & = & \tph \left( \sum_{k=0}^{\nph-1} v_k + \frac{1}{2} v_{\nph} \right)\,.
\end{eqnarray}

Inserting Eq.\ \ref{eq:vk} and solving for \tph\ results in

\begin{equation}\label{eq:taufirst}
 \tph = \frac{L \sqrt{\meff/2}}{\sum_{k=0}^{\nph -1} \sqrt{\ekin^0 - k \eph} + \frac{1}{2} \sqrt{\ekin^0 - \nph
\eph}}\,.
\end{equation}

We estimate the distance the electrons travel to reach the collector barrier (by going along the edge) to be about
$L(B) \simeq1.5\,\mu$m [compare Fig.\ \ref{sample}(a)] for the magnetic field values considered here; for $B = 0$ no
transport along the edge is expected and $L = L_{\text{EC}}$ is used instead.

\begin{figure}[ht]
\includegraphics[width=\columnwidth]{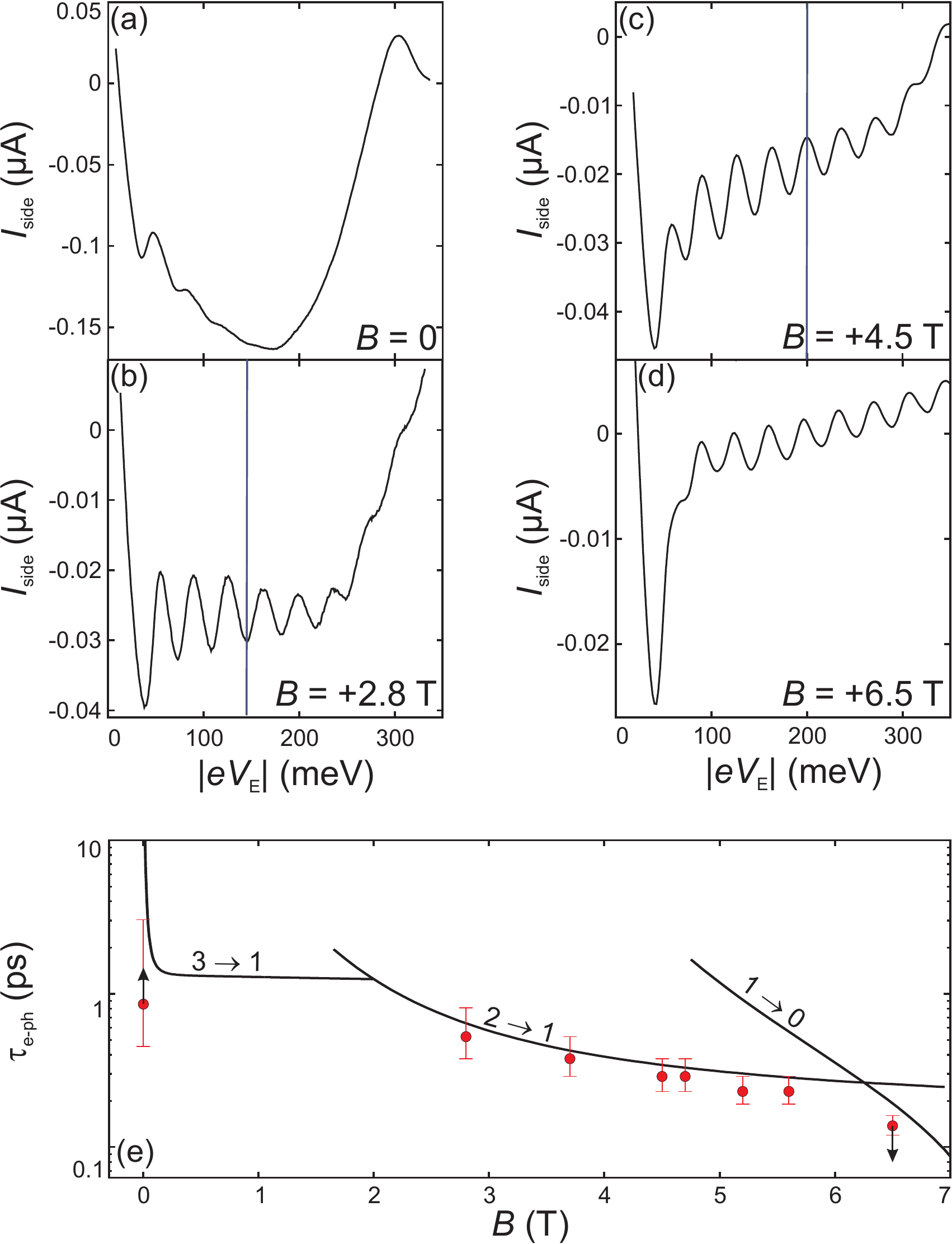}%
\caption{\label{differentfields}(Color online) \iside\ as a function of \eve\ at constant $\iE = \unit{100}{\nano\ampere}$  as in Fig.\
\ref{current_color}(b) and \ref{current_color}(c) for various perpendicular magnetic fields. The vertical lines in (b)
and (c) mark the value of \eve\ above which the amplitude of the phonon-induced oscillations starts to decrease
considerably. For (a) and (c) these values of \eve\ are outside of the plots. (e) Scattering times displayed in Table
\ref{table} as a function of magnetic field (determined from 
(a)--(d) and additional measurements). The leftmost (rightmost) data point represents a lower (upper) limit as indicated
by arrows (see Table \ref{table}). The solid lines show theory
curves obtained by heuristic application of Bockelmann's theory (numerical calculations, no fit
parameters involved). The three branches belong to transitions between electronic subbands ($3 \to 1$, $2 \to 1$, $1
\to 0$) in $x$ direction via the emission of an optical phonon. }
\end{figure}

Fig.\ \ref{differentfields}
shows the measured \iside\ as a function of \vE\ for several magnetic fields at an injected current $\iE =
\unit{100}{\nano\ampere}$. Assuming that the emission of optical phonons is indeed by far the most efficient scattering
process for hot electrons with $\ekin > \eF + \eph$ we expect a constant oscillation amplitude of \iside\ as long as
$ \nph^\mathrm{max} < \nph(L)$. This means that all optical phonons that are energetically allowed are emitted before
the electron reaches the collector barrier. Accordingly we can interpret an onset of a decrease of the oscillation
amplitude of \iside\ as a function of $\left| e \vE \right|$ as the electron energy at which $\nph(L) \simeq 
\nph^\mathrm{max}$. This energy is marked in Figs.\ \ref{differentfields}(b) and \ref{differentfields}(c) by vertical
lines and corresponds to $\ekin^0 = \eF + \nph^\mathrm{max} \eph = \eF + \nph(L) \, \eph$. Inserting this
relation into Eq.\ \ref{eq:taufirst} then yields

\begin{equation}
\label{eq:tausecond}
\tph =\frac{L \sqrt{\meff/2}}{\sum_{k=1}^{\nph} \sqrt{\eF +  k \eph} + \frac{1}{2} \sqrt{\eF}}
\end{equation}

which can be used to calculate \tph\ from the data for different magnetic fields.

For $B = 0$ [Fig.\ \ref{differentfields}(a)] the phonon-induced oscillations quickly weaken
as the excess energy \eve\ is increased. We interpret the absence of a constant oscillation amplitude as a scattering time
that exceeds the calculated value for $\nph(L)=1$. In addition, for $B=0$ our assumption of forward scattering
only does not
hold, which tends to increase the effective $L$. Hence, our model tends to underestimate \tph\ for $B=0$. For $B =
\unit{6.5}{\tesla}$ [Fig.\
\ref{differentfields}(d)], the oscillation amplitude is constant over the whole energy range measured, showing 9
oscillations, so
we conclude $\nph(L) \ge9$. The values for $\nph(L)$ and $\tph$ obtained from the data for different magnetic
fields shown in Figs.\ \ref{differentfields}(a) -- \ref{differentfields}(d) and similar measurements are summarized in
Table \ref{table}.
\begin{table}[ht]
\begin{tabular}{c|c|c|c|c|c|c|c|c}
$B$ (T)    & 0            &2.8       &3.7       &4.5 	&4.7      &5.2       &5.6       &6.5\\\hline
$\nph$     & $\lesssim1$  &3    &4  &5  &5	  &6  &6       &$\gtrsim9$\\\hline
\tph\ (ps) & $\gtrsim0.85$&$0.52$&$0.38$&$0.29$&$0.29$&$0.23$&$0.23$&$\lesssim0.14$
\end{tabular}
\caption{\label{table}Estimated number of optical phonons $\nph(L)$ emitted within the distance $L$ that hot
electrons travel between
emitter and collector (for the algorithm  of determining $\nph(L)$ see Fig.\ \ref{differentfields}). The
electron-phonon
scattering
times \tph\ are then calculated with Eq.\ (\ref{eq:tausecond}).}
\end{table}
In Fig.\ \ref{differentfields}(e) \tph\ (logarithmic scale) is plotted as a function of the magnetic field. The
errorbars express the uncertainty in determining $\nph(L)$ taken to be $\Delta\nph=1$. For the two extremal values, $B =
0$ and $B = \unit{6.5}{\tesla}$, the plotted values of \tph\ represent a lower ($B = 0$) and upper ($B =
\unit{6.5}{\tesla}$) limit, respectively.

In summary we observe a dramatic decrease of \tph\ by more than an order of magnitude as the perpendicular magnetic
field is increased from zero to a filling fraction beyond $n=1$.

\subsection{Theoretical analysis of electron-phonon scattering}

The interaction between hot electrons and optical phonons  in nanostructures has been
investigated theoretically in detail with and without applied perpendicular magnetic field.\cite{Sarm2,bockelmann_theory,Gass,tel,bru}

Extending the calculations of Bockelmann \emph{et al.}\cite{bockelmann_theory}\ Telang \emph{et al.}\cite{tel}\ investigated
the effect of a magnetic field on the electron-phonon scattering rate in quasi--one-dimensional systems rigorously by
solving the Schr\"odinger equation with boundaries. They found that the scattering rate only weakly depends on the electron energy
in a quasi one-dimensional channel. Relying on this weak energy dependence, for convenience here we numerically
integrate Bockelmann's formula for a fixed
initial energy of $\ekin=82\,$meV of the injected electron. This energy value has been chosen to be similar to the one used
in Ref.\ \onlinecite{heiblum_phonon} where the electron-phonon scattering time in a vertical tunneling structure is calculated
 for zero magnetic field (their result of $\lph=0.24\,$ps is not applicable to our case, though, since we have a
lateral
nanostructure in a 2DES). 

We use the edge channel diameter $\rc =\sqrt{2(n+0.5)}\,\lc$ as the lateral confinement
width. The $x$ component of the phonon momenta (along the direction of electron confinement) can be treated as a
continuum since $L_x \gg \rc$. We assume the electrons to occupy the lowest perpendicular subband $k_z\rightarrow 0$
(compare last paragraph of Sec.\ \ref{sec-review}), hence phonon emission mainly occurs within the plane of the 2DES.
The solid lines in Fig.\ \ref{differentfields}(e) express our numerical results for $\tph(B)$. They are in excellent
agreement with experimental data even though they contain no free parameters. These theory curves correspond to three
branches of intersubband scattering (subbands in $x$ direction) under emission of an optical phonon. Most experimental
data points lie just below the lowest branch which is expected if the overall scattering rate is dominated
by the fastest process [lowest branch Fig.\ \ref{differentfields}(e)] with a slight contribution of slower
processes (higher branches). The transitions between the three branches result from the according discontinuities in the
electronic one-dimensional density of states (see Bockelmann's results\cite{bockelmann_theory}). These jumps have been
observed and discussed in detail in numerical calculations by Telang \emph{et al.}\cite{tel}.

\section{Influence of the magnetic field direction}
\label{sec-edges}

\begin{figure}[ht]
\includegraphics[width=\columnwidth]{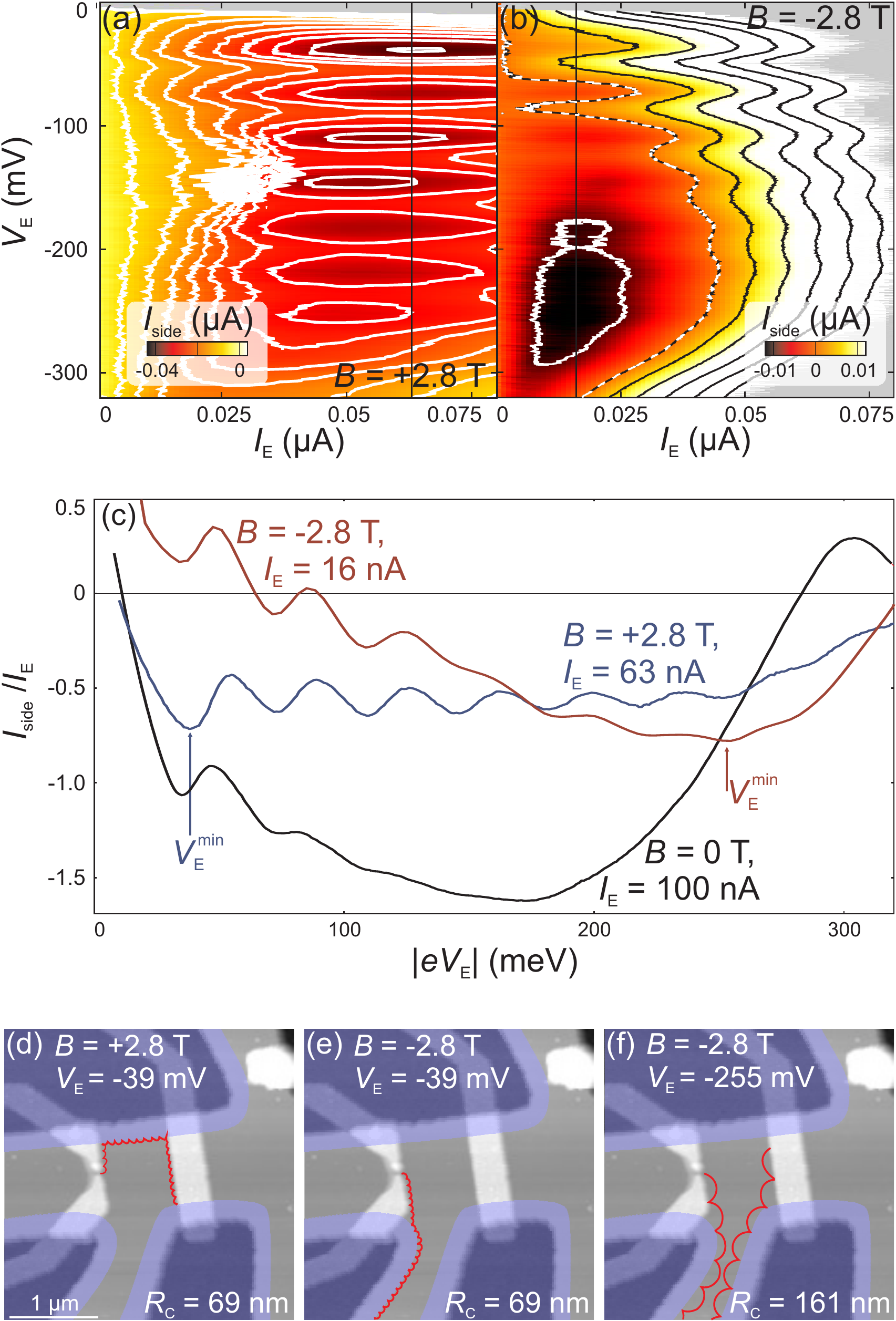}%
\caption{\label{cyclotron}(Color online) (a,b) \iside\ as a function of \iE\ and \vE\ as in Fig.
\ref{current_color}, for $\vBC = \unit{-280}{\milli\volt}$ $(\eBC = \eF -
\unit{1.4}{\milli\evolt}$) at perpendicular magnetic fields of $B=2.8\,$T and $B=-2.8\,$T. Contour lines are spaced by
\unit{5}{\nano\ampere}. (c) Vertical traces ($\iside/\iE$ is plotted as a function of \eve) through the absolute minimum
of \iside\ for three different magnetic fields:  $B=2.8\,$T at $\iE=16\,$nA (a), $B=-2.8\,$T at $\iE=63\,$nA (b) and
$B=0$ at $\iE=100\,$nA (from data similar to Fig.\ \ref{current_color}(a) but for $\vBC =
\unit{-280}{\milli\volt}$ as in \ref{cyclotron}(a) and (b) ). (d--f)
Sketches of the classically expected skipping orbit motion of hot electrons with energies of $\eve=39\,$meV (d,e) and
$\eve=255\,$meV (f) corresponding to the absolute minima of \iside\ at $B=\pm 2.8\,$T. A zone of depletion extending an
estimated \unit{200}{\nano\meter} from the etched edges into the Hall bar is marked in blue.}
\end{figure}

Until now, only perpendicular magnetic fields that guide the injected electrons away from the side contact have been discussed
($B>0$). In this section, the discussion is extended to the opposite field direction ($B<0$) that tends to direct the injected
electrons into the side contact [compare Figs.\ \ref{cyclotron}(d) and \ref{cyclotron}(e)]. In a na\"{\i}ve model one
would expect
$\iside\equiv\iE$ (and $\iC=0$) for $B <0$ as long as $\left| B \right|$ is large enough to confine the current to edge channels,
and for a perfect side contact without reflection. Figs.\ \ref{cyclotron}(a) and \ref{cyclotron}(b)
show measurements of \iside\ as a function of \vE\ and (relatively small) \iE\ for opposite field directions $B =
\unit{$\pm$2.8}{\tesla}$. All other
parameters are kept equal (the color scale has been adapted, though). As in the previous figures,
white contour lines denote $\iside<0$ while for black lines  $\iside>0$. Vertical traces (for constant \iE) including the
respective absolute minimum of \iside\ are plotted in Fig.\ \ref{cyclotron}(c) for both magnetic field directions $B =
\unit{$\pm$2.8}{\tesla}$ as well as for $B=0$.

\begin{figure}[ht]
\includegraphics[width=\columnwidth]{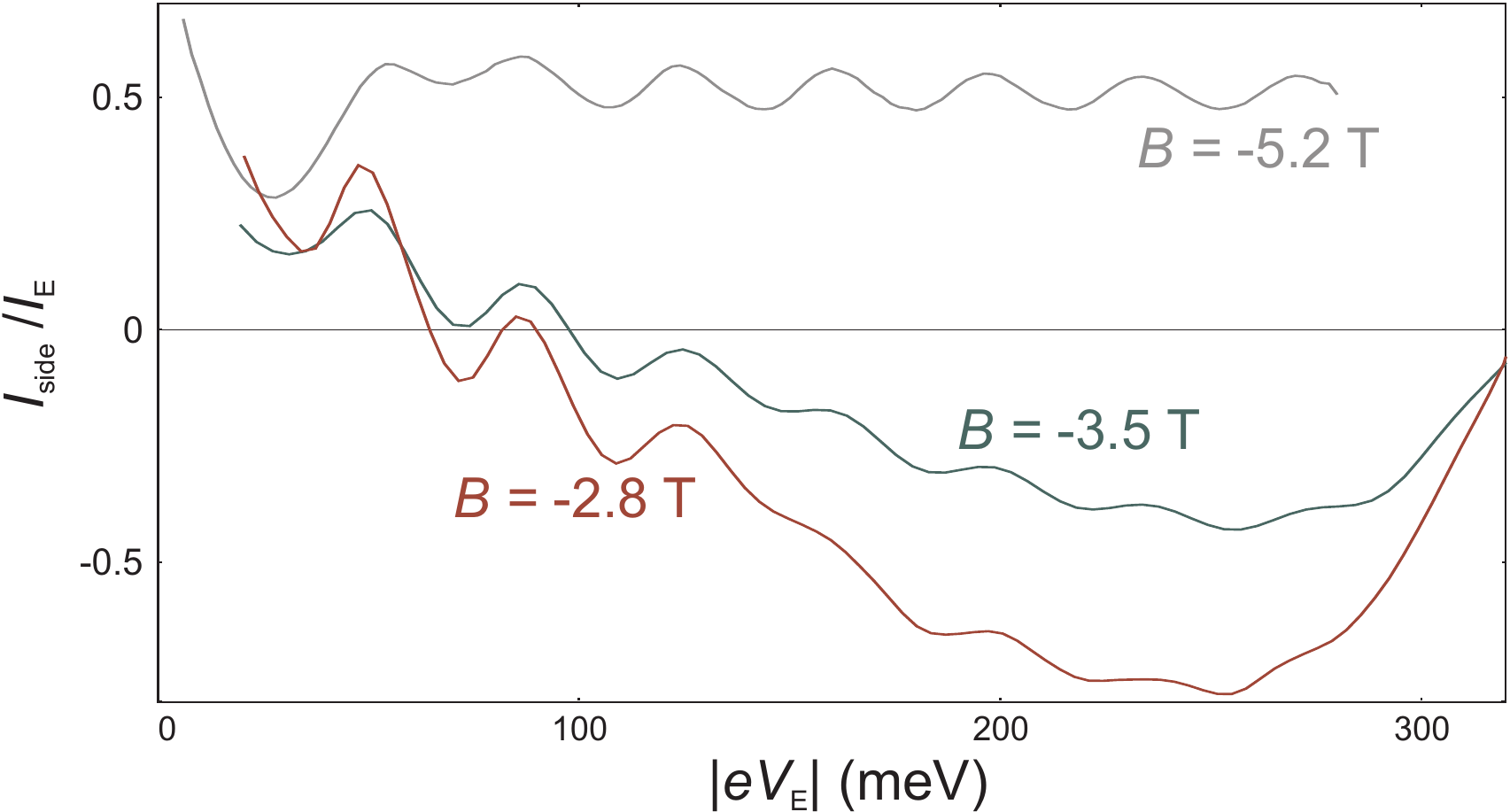}%
\caption{\label{otherfields}(Color online) $\iside/\iE$ versus \eve\ for constant $\iE=15\,$nA. Included is a
measurement for $B=-2.8\,$T as already shown in Fig.\ \ref{cyclotron}(c) and two additional curves for $B=-3.5\,$T and
$B=-5.2\,$T measured at $\vBC = \unit{-265}{\milli\volt}$ $(\eBC = \eF - \unit{1.8}{\milli\evolt}$. The slightly
different collector barrier height for $B=-2.8\,$T of $\eBC = \eF - \unit{1.4}{\milli\evolt}$ is not important for the
qualitative behavior of interest here.}
\end{figure}

The na\"{i}ve expectation $\iside=\iE$ for $B<0$ based on edge channel transport is clearly not fulfilled. Instead we observe both
oscillations of \iside\ induced by the emission of optical phonons and amplification ($\iside<0$). Nevertheless, the
differences for opposite magnetic field directions are striking. For $B<0$ the phonon-induced oscillations decrease much
quicker as \eve\ is
increased and the region of amplification ($\iside<0$) is much smaller. In addition the maximum amplification effect
[absolute
mimimum of \iside\ marked by $\vE^\mathrm{min}$ in Fig.\ \ref{cyclotron}(c)] occurs at a much larger energy
$|e\vE^\mathrm{min}|\simeq250\,$meV for $B<0$ compared to $|e\vE^\mathrm{min}|\simeq35\,$meV for $B>0$. The absolute minimum for
$B = 0$ occurs between these two values at $|e\vE^\mathrm{min}|\simeq180\,$meV. For even more negative magnetic fields, \iside\
stays positive for all \eve, thus no amplification effect can be observed at all. This can be seen in Fig.\
\ref{otherfields}
which shows the $B = \unit{-2.8}{\tesla}$ trace from Fig.\ \ref{cyclotron}(c) as well as measurements for $B =
\unit{-3.5}{\tesla}$ and \unit{-5.2}{\tesla}. The latter field is large enough to prevent amplification ($\iside>0$).

The Landauer-B\"uttiker edge channel picture is limited to linear-response transport for $|\ekin-\eF|\ll\eF$, and is therefore not
appropriate to describe our data at $\eve\gg\eF$. In this non-linear regime scattering of the hot electrons, incomplete screening
of the Lorentz force acting on hot electrons in the bulk of the 2DES, and the energy-dependent cyclotron radius have to be
accounted for. In the following, we develop a semiclassical approach to model the magnetic field dependence far from equilibrium.
As before we assume that the cyclotron radius $\rc=\frac{\sqrt{2\meff\ekin}}{|eB|}$ roughly defines the width of the channel in
which most of the hot electrons move along the edge of the 2DES. This is assured by the large Lorentz force acting on the hot
electrons which forces them back to the edge even after scattering.

The cyclotron radius at the energy $|e\vE^\mathrm{min}|\simeq35\,$meV of the absolute minimum of \iside\ for $B=+2.8\,$T is
$\rc\simeq69\,$nm, still relatively small. Hence the electrons move along the edge in a narrow channel as sketched in Fig.\
\ref{cyclotron}(d). For the reversed magnetic field $B=-2.8\,$T electrons are guided towards the side contact as
shown in Fig.\
\ref{cyclotron}(e). This should lead to $\iside=\iE$, and indeed \iside\ becomes positive for relatively small
$\eve\lesssim100\,$meV, where we observe the maximum amplification for $B>0$ [Fig.\ \ref{cyclotron}(c)]. Fig.\
\ref{cyclotron}(f)
sketches the situation at an energy of $|e\vE^\mathrm{min}|\simeq250\,$meV, corresponding to $\rc=161\,$nm at $B =
\unit{-2.8}{\tesla}$. At this energy, $2\rc$ is already comparable to the width of the side contact and energy exchange
between the two edge channels with opposite current direction can occur. As a result excited electrons can reach the collector
barrier and amplification can occur similarly to $B>0$. Furthermore, within the region of backscattering (at the
``mouth'' of the side
contact) electron-electron scattering is enhanced as the two ``edge channels'' of opposite direction tend to merge into a region
without a preferred direction. For large energies \eve\ the side current therefore becomes more negative for $B<0$ compared to
$B>0$.

Our semiclassical model based on edge transport of hot electrons in a strong perpendicular magnetic field fails to explain one crucial
observation, though. In the region of relatively small energies $\eve\lesssim100\,$meV where $\iside>0$ for $B<0$ we observe
phonon-induced oscillations while our model does \emph{not} propose such an effect since it
assumes that all current-carrying electrons are guided towards the side contact. In this case the emission of optical phonons
should not influence the number of hot electrons reaching this contact. Even backscattering at a non-perfect ohmic
contact --- which
is in our case one millimeter away --- would not allow such a behavior since after having traveled such a distance all
electron-hole
pairs would have recombined within the edge channel, and no effects caused by phonon emission should be visible. In addition we
would expect to see $\iside=\iE$, but Fig.\ \ref{cyclotron}(c) shows that this is not the case. The observation of
phonon-induced
oscillations in this regime points towards leakage of energy into bulk states related to the emission of optical phonons by hot
electrons moving along the edge. However, free hot electrons scattered into the bulk of the 2DES tend to be directed back to the
edge of the 2DES by the Lorentz force acting on them. This contradiction can be resolved by taking localized bulk states
into account which are a result of disorder combined with reduced screening in a magnetic field. A hot electron can
travel from BE to BC by a combination of energy relaxation and hopping between these localized states. Our experimental
observations are consistent with the assumption that
in a large magnetic field a small part of the overall current is carried by such a hopping
transport mechanism. Finally, we observe that the amplitude of the phonon-induced oscillations decays faster as the
energy \eve\ is increased
for $B<0$
compared to $B>0$. This behavior might be related to the transport via localized states but will not be discussed in detail here.

\section{Conclusions}

In conclusion we have studied the energy relaxation of hot electrons injected at an energy $\ekin\gg\eF$ into an otherwise
degenerate 2DES. The transport measurements have been performed in a mesoscopic three-terminal device in which two of the contacts
(emitter and collector) are separated by tunable electrostatic barriers. The emitter current and the energy of the injected
electrons are fully controlled while the currents into the other two contacts are measured. Our main observation is that as a
function of a perpendicular magnetic field the electron-electron scattering rate decreases while the emission of optical phonons
increases drastically. Quantitatively, the corresponding electron-phonon relaxation time declines from over 2\,ps at zero magnetic
field to below 0.2\,ps at strong perpendicular magnetic fields. Numerical calculations within the theoretical standard approaches quantitatively
confirm our data and support the following interpretation of our experiments. We have observed the transition from two-dimensional
scattering of hot electrons at zero magnetic field towards one-dimensional dynamics at large perpendicular magnetic
fields. In
the quasi--one-dimensional limit the interaction of hot electrons with optical phonons becomes so strong that we clearly
observe the emission of more than 10 optical phonons by individual hot electrons in the current signals. This allows us to
determine the onset energy of LO phonons in GaAs at $T\simeq260\,$mK with high precision to $\eph=36.0\pm0.1\,$meV.
Finally,
measurements as a function of the direction of a strong perpendicular magnetic field suggest that the expected flow of
the hot electrons along
the mesa edges is accompanied by a second transport contribution which we interpret as hot electrons hopping between
localized
bulk states of the 2DES.

\begin{acknowledgments}

We thank J.\,P.\ Kotthaus and A.\ Govorov for helpful discussions. Financial support by the German Science Foundation
via SFB 631, LU 819/4-1, and the German Israel program DIP, the German Excellence Initiative via the "Nanosystems
Initiative Munich
(NIM)", and LMUinnovativ (FuNS) is gratefully acknowledged.
\end{acknowledgments}

\appendix*
\section{Current dependence}

\begin{figure}[ht]
\includegraphics[width=\columnwidth]{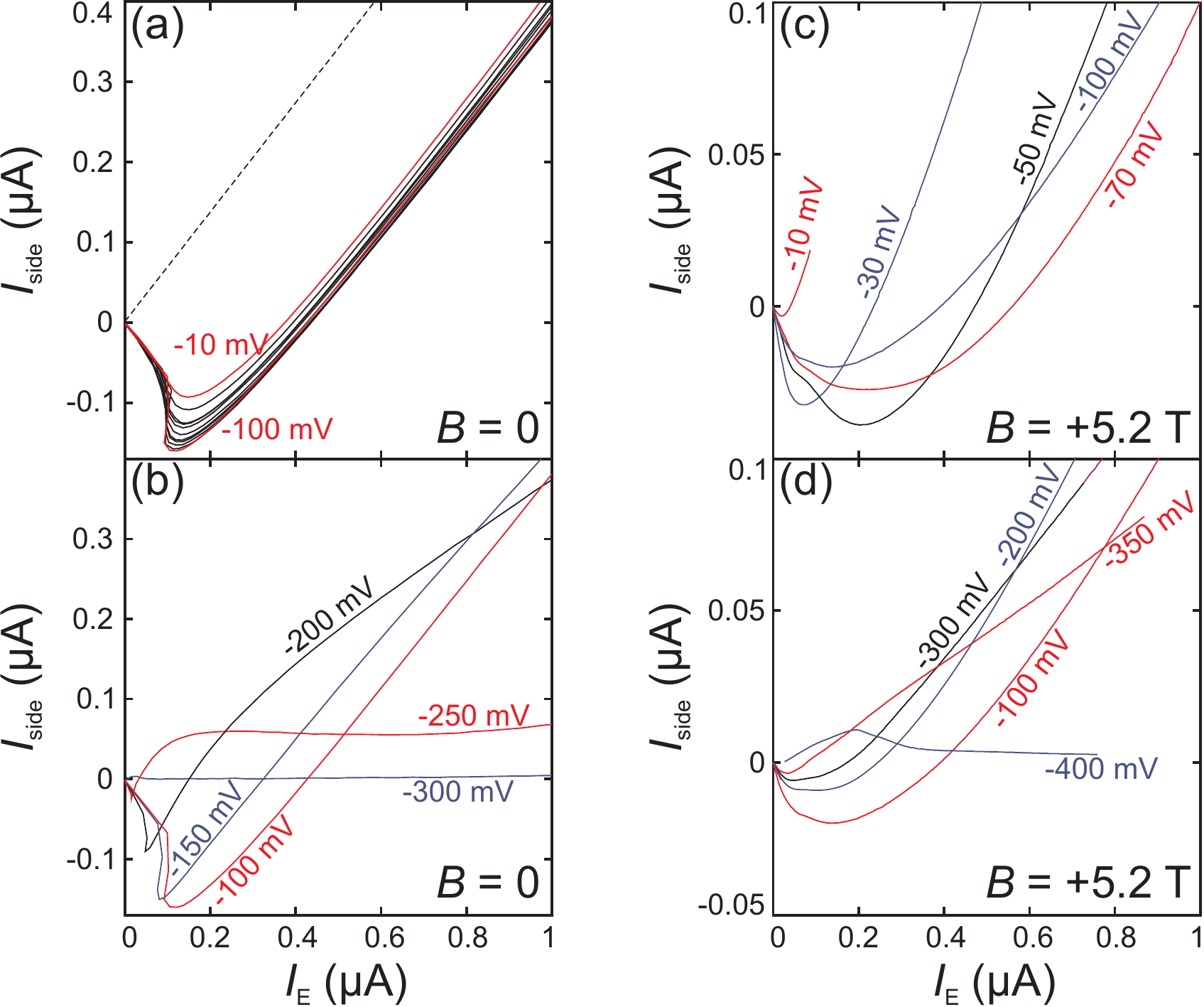}%
\caption{\label{current_line} (Color online) \iside\ as a function of \iE\ for several values of $\vE$ (given
in the
plot); $\vBC = \unit{-265}{\milli\volt}$ $(\eBC = \eF -\unit{1.8}{\milli\evolt}$. $B=0$ in (a,b) and $B=5.2\,$T in
(c,d). Curves for relative low excess energies $10\,\mathrm{meV}\le\eve\le100\,$meV [equal spacings of 10\,meV in (a)
and energies as marked in (c)] are plotted in (a,c) while the plots in (b,d) concentrate on $\eve\ge100\,$meV (energies
marked).}
\end{figure}

The main focus of this article is to study \iside\ as a function of the kinetic excess energy \eve\ of the injected
electrons.
Fig.\ \ref{current_line}
shows the side current \iside\ as a function of \iE\ for $B=0$ (left) as well as for $B=5.2\,$T (right). The $B=0$
data have already been discussed in Ref.\ \onlinecite{hotelectrons}. Data for $\eve\le100\,$meV are plotted in Fig.\
\ref{current_line}(a). In this low-energy regime, \iside\ first rapidly decreases as a function of \iE\ before it grows
again for
larger currents. Strikingly, for $\iside>200\,$nm all curves measured at energies below $\eve=100\,$meV follow the
same straight line which is actually parallel to the line expected for ohmic behavior [dashed in Fig.\
\ref{current_line}(a)].\cite{hotelectrons}. This curve
shape can be interpreted as an ohmic contribution to \iside\ (which is proportional to \iE\ and is determined by the
ohmic
resistances of the three-terminal device) plus a negative contribution that saturates at $\iside\simeq-0.2\,\mu$A. The
saturation of the negative contribution to \iside\ is explained by taking into account the neutralization of electron-hole
pairs created by electron-electron scattering. The generation rate of the electron-hole pairs should be roughly proportional to
the number of injected electrons, so neutralization becomes more efficient at larger \iE. This is related to the positive charge
building up between BE and BC due to the amplification effect which hinders the escape of hot electrons via BC. Since electrons
that cannot escape add to the neutralization of holes a steady state is reached in which the negative contibution to \iside\
saturates.\cite{hotelectrons} 

The ohmic contribution of \iside\ proportional to \iE\ is straightforward to explain. At energies at which the electron-electron
scattering length \lee\ is small, multiple scattering processes result in many electrons with small kinetic energies and almost
arbitrary direction of their momentum. These electrons mimic a diffusive motion, the prerequisite for an ohmic
behavior.\cite{hotelectrons}

The regime of high energies $\eve\ge100\,$meV is depicted in Fig.\ \ref{current_line}(b). Here we observe the transition
to \lee\
being longer than the device dimensions, which | in its extremes | results in $\iside=0$ (observed at $\eve=300\,$meV) because
the hot electrons move ballistically through the sample.\cite{hotelectrons}

In a strong perpendicular magnetic field the amplification effect is reduced, hence Fig.\ \ref{current_line}(c) has a
differently
scaled y-axis. The preferred directional motion along the edges of the Hall bar at $B=5.2\,$T prevents ohmic contributions even at
large currents. This explains part of the more complex behavior seen for low energies $\eve\le100\,$meV in Fig.\
\ref{current_line}(c) at $B=5.2\,$T, namely that the bunching of curves \iside(\iE) for several energies at high \iE\ is
missing.
In addition, in a large positive magnetic field the minimum of \iside\ strongly shifts as a function of energy and $\iside<0$
persists to larger \iE\ and larger energies \eve\ compared to $B=0$ [compare Figs.\ \ref{current_line}(c) and
\ref{current_line}(d)].
A detailed discussion of these effects which reflect the more complex situation in a strong perpendicular magnetic field including the
enhanced emission of optical phonons would be very difficult. In a positive magnetic field the high-energy limit $\iside=0$ is
reached at a larger
\eve\ as for $B=0$ [see also Figs.\ \ref{current_color_zerofield}(c) and \ref{current_color}(a)]. This tendency might be
understood in terms of the stronger interaction with optical phonons which shifts the regime of ballistic motion
throughout the Hall bar towards higher velocities of the injected electrons.

% \bibliography{opticalphonons}

\end{document}